\documentclass[aps,prd,twocolumn,superscriptaddress,
showpacs,preprintnumbers,nofootinbib,10pt]{revtex4-1}
\usepackage{amsmath,amssymb,hyperref,subfigure,xspace}
\usepackage{mciteplus,color,mathtools,graphicx}
\usepackage{multirow}
\usepackage[english]{babel}
\hypersetup{pdfnewwindow=true,colorlinks=true,
linkcolor=blue,citecolor=blue,filecolor=blue,
urlcolor=blue 
} 
\newcommand{\ceem}{Center for  Exploration  of  Energy  and  Matter,  
Indiana  University,  
Bloomington,  IN  47403,  USA}
\newcommand{\ghent}{Department of Physics and Astronomy, 
Ghent University, Ghent 9000, Belgium}
\newcommand{\icn}{Instituto de Ciencias Nucleares, 
Universidad Nacional Aut\'onoma de M\'exico, 
Ciudad de M\'exico 04510, Mexico}
\newcommand{\ectstar}{European Centre for Theoretical Studies in Nuclear Physics and Related
Areas (ECT$^*$) and Fondazione Bruno Kessler,
I-38123 Villazzano (TN), Italy}
\newcommand{\indiana}{Physics  Department,  
Indiana  University,  
Bloomington,  IN  47405,  USA}
\newcommand{\indianaInf}{School of Informatics and Computing, Indiana University, Bloomington, IN 47405, USA}
\newcommand{\jlab}{Theory Center,
Thomas  Jefferson  National  Accelerator  Facility, 
Newport  News,  VA  23606,  USA}
\newcommand{\mainz}{Institut f\"ur Kernphysik \& PRISMA Cluster of Excellence, 
Johannes Gutenberg Universit\"at, 
D-55099 Mainz, Germany}
\newcommand{\murcia}{Departamento de F\'isica, 
Universidad de Murcia, 
E-30071 Murcia, Spain}
\newcommand{\jpac}{Joint Physics Analysis Center}
\begin{document}
\title{Regge phenomenology of the $N^*$ and $\Delta^*$ poles}
\author{J.~A.~Silva-Castro}
\email{jorge.silva@correo.nucleares.unam.mx}
\affiliation{\icn}
\author{C.~Fern\'andez-Ram\'{\i}rez}
\email{cesar.fernandez@nucleares.unam.mx}
\affiliation{\icn}
\author{M.~Albaladejo}
\affiliation{\jlab}\affiliation{\murcia}
\author{I.~V.~Danilkin}
\affiliation{\mainz}
\author{A.~Jackura}
\affiliation{\ceem}\affiliation{\indiana}
\author{V.~Mathieu}
\affiliation{\jlab}
\author{J.~Nys}
\affiliation{\ghent}
\author{A.~Pilloni}
\affiliation{\jlab}\affiliation{\ectstar}
\author{A.~P.~Szczepaniak}
\affiliation{\jlab}\affiliation{\ceem}\affiliation{\indiana}
\author{G.~Fox}
\affiliation{\indianaInf} 
\collaboration{\jpac}
\noaffiliation
\preprint{JLAB-THY-18-2797}
\begin{abstract}
We use Regge phenomenology to 
study the structure of the poles of the 
$N^*$ and $\Delta^*$ spectrum.
We employ the
available pole extractions from partial wave analysis of meson
scattering and photoproduction data.
We assess the importance of the imaginary 
part of the poles (widths) to obtain a consistent 
determination of the parameters of the Regge trajectory.
We compare the several pole extractions and
show how Regge phenomenology can be used to 
gain insight in the internal structure of baryons.
We find that the majority of the states in the parent Regge trajectories 
are compatible with 
a mostly compact three-quark state picture.
\end{abstract}
\maketitle
\section{Introduction}
\label{sec:introduction} 
The baryon spectrum is  
one of the main tools for  investigation of the  
nonperturbative QCD phenomena. 
In particular, the low-lying non-strange sector 
containing the $N^*$ and $\Delta^*$ resonances, 
which is accessible 
in pion-nucleon  scattering and photoproduction
experiments, is a primary source of insights into the quark model.  
The goal of baryon spectroscopy is to understand the origin and structure 
of resonances, {\it e.g.} to establish if a given resonance 
can be classified as compact three quark ($3q$) state, 
as predicted by the quark model 
or that it has other hadronic components. 
This is often done through partial wave analyses, with 
resonances appearing in  individual partial waves  
that are independently parametrized to fit the data. 
Such analyses miss  global constraints imposed 
by the Regge theory that 
connect partial waves through 
analyticity in the angular momentum 
plane~\cite{Collins:1977jy,Gribov:2003nw,Gribov:2009zz}.
According to Regge theory,
resonances appear as poles in the  
angular momentum plane. 
The pole location, which changes as a function 
of the resonance mass and defines the 
so-called Regge trajectory,  
can be used to study the microscopic mechanisms 
responsible for resonance 
formation~\cite{Londergan:2013dza,Carrasco:2015fva,Fernandez-Ramirez:2015fbq,Pelaez:2017sit}.

The most noticeable feature of the hadron spectrum is that its 
Regge trajectories are approximately 
linear. This was first shown by 
 Chew and Frautschi 
 ~\cite{Chew:1962eu} who plotted spin of  resonances $J_p$ vs their mass squared $M^2$, which, in the narrow width approximation 
  corresponds to a Regge trajectory.  The patterns implied by the Chew-Frautschi plot 
can be used
to guide partial wave analyses. 
For example, gaps in the trajectories hint to missing states.  
The approximate linearity of Regge trajectories 
is one the strongest phenomenological indications of 
confinement~\cite{Greensite:2011zz} 
and therefore states belonging to
linear trajectories are expected 
to be closely connected to quark model    
predictions~\cite{Rossi:1977cy,Montanet:1980te}.

Resonance decays, contribute to  trajectories 
by introducing imaginary parts. 
These are constrained by unitarity and
analyticity, and are related  to resonance widths~\cite{Gribov:1962fx}.
Consequently, Regge trajectories are a mapping of the 
complex energy plane, the $s$-plane, onto the complex 
angular momentum, 
the $J$-plane. 
More specifically, 
 since a  resonance is characterized by its  complex energy $s_p$ and spin $J_p$, Regge
  trajectory $\alpha(s)$ is a complex function
 such that $\alpha(s_p) \equiv  
 (\Re (J(s_p)), \Im(J(s_p)) )=
 (J_p,0)$.
 \footnote{
 The symbols $\Re$ and $\Im$ stand for the
 real and imaginary parts, respectively.} 
Hence, in the general case of finite resonance widths the Chew-Frautschi plot has to be interpreted 
 as the relation between 
  $\Re(s_p)$ {\it vs.} $\Re(J)=J_p$.
We note that, as we are no longer using the narrow width approximation,
the Chew-Frautschi plot no longer provides a complete description of the Regge trajectory and 
when analyzing  twodimensional plots one can consider additional relations, like 
$\Im(s_p)$ {\it vs.} $\Re(J)=J_p$ ~\cite{Fernandez-Ramirez:2015fbq},
to fully characterize the Regge trajectory,
or surface plots of $\Re(\alpha(s))$ as a function of  complex $s$, however, will continue referring to the  Chew-Frautschi plot as mapping of real mass onto real spin.

In the past, resonance poles were often not computed and,  
with a few exceptions~\cite{Fiore:2000fp,Fiore:2004xb}, 
fits to the Chew-Frautschi plots 
gave the only information about the Regge trajectory.
Constituent quark model predictions for hadron masses 
adhere nicely to the approximately linear behavior both in the
baryon~\cite{Nakkagawa:1972az,Bijker:1994yr,Bijker:2000gq,Ortiz-Pacheco:2018ccl,Capstick:1986bm,Loring:2001kx,Inopin:1999nf,Tang:2000tb} 
and the
meson~\cite{Godfrey:1985xj,Koll:2000ke,Tang:2000tb,Ebert:2009ub} 
sectors.
Flux tube models of baryons
also provide linear trajectories~\cite{Isgur:1984bm,Olsson:1995mv,Semay:2007cv}.

In this article,
following the analysis of the strange baryon
sector~\cite{Fernandez-Ramirez:2015fbq} 
we use Regge phenomenology 
to study the $N^*$ and $\Delta^*$ spectra. 
Resonance  pole masses and widths are nowadays more prominently 
featured in the Particle Data Group (PDG) 
tables~\cite{Tanabashi:2018oca}. 
This is because, in the last years,  
amplitude analyses have become more 
sophisticated enabling for extraction of resonance  poles from the experimental data. 
We fit complex Regge trajectories  to the  spectra obtained by several partial wave
analyses~\cite{Cutkosky:1979fy,Cutkosky:1980rh,Ronchen:2018ury,Anisovich:2011fc,Sokhoyan:2015fra,Svarc:2014aga,Svarc:2014zja}
of meson scattering and photoproduction data. The objectives of this article are:
(i) to provide a comprehensive comparison 
of the different $N^*$ and $\Delta^*$ pole extractions based on Regge phenomenology;
(ii) to assess the impact of neglecting 
the imaginary part of the poles in the computation of the 
Regge trajectory, in particular in the extraction of the slope 
parameter that can be compared to the one used 
in fits to the high energy 
proton-antiproton data~\cite{VandeWiele:2010kz}; 
and (iii) to guide future $N^*$ and $\Delta^*$ pole
extractions~\cite{Nys:2016vjz,Mathieu:2017but,Mathieu:2018mjw}.
The paper is organized as follows.
In Sec.~\ref{sec:poles} we review the 
$N^*$ and $\Delta^*$ spectra available in the literature 
that will be used in our analysis.
In Sec.~\ref{sec:models} we describe the phenomenological 
models used to fit the spectrum and in Sec.~\ref{sec:results} 
we explain the fitting procedure, present the results and 
discuss the statistical analysis. 
Conclusions are given in Sec.~\ref{sec:conclusions}. 
\section{$N^*$ and $\Delta^*$ pole extractions}
\label{sec:poles}
\begin{table*}
\caption{Summary of pole positions $M_p,\Gamma_p$ in MeV
for $I^\eta=\frac{1}{2}^+$ states
where $M_p=\Re\, \left[\sqrt{s_p}\right]$ 
and $\Gamma_p=-2 \, \Im\, \left[ \sqrt{s_p} \right]$.
$I$ stands for isospin, $\eta$ for naturality, 
$J_p$ for spin, and 
$P$ for parity.
Naturality and parity are related by 
$\eta=\tau P$ where $\tau$ is the signature.
For baryons, $\eta=+1$, natural parity, 
if $P=(-1)^{J_p-1/2}$, and
$\eta=-1$, unnatural parity, if $P=-(-1)^{J_p-1/2}$.}
\label{tab:regge12p}
\begin{ruledtabular}
\begin{tabular}{c|ccccc}
Name & $N(939)$ & $N(1520)$ & $N(1680)$& $N(2190)$&
$N(2220)$\\
Status & **** & **** & **** & **** & ****\\
$I^\eta_{(\tau)}\, J^P_p$& 
$\frac{1}{2}^+_{(+)}\, 1/2^+$ &
$\frac{1}{2}^+_{(-)}\, 3/2^-$ &
$\frac{1}{2}^+_{(+)}\, 5/2^+$ &
$\frac{1}{2}^+_{(-)}\, 7/2^-$ &
$\frac{1}{2}^+_{(+)}\, 9/2^+$ \\
\hline
CMB    
&939(1),\, 0&1510(5),\, 114(10) & 1667(5),\, 110(10)& 2100(50),\, 400(160)& 2160(80),\, 480(100)\\
J\"uBo 
&939(1),\, 0&1509(5),\, \phantom{0}98(3) & 1666(4),\, \phantom{0}81(2) & 2084(7),\, 281(6)&2207(89),\, 659(140) \\
BnGa 
&939(1),\, 0&1507(3),\, 111(5) & 1676(6),\, 113(4) & 2150(25),\, 325(25) & 2150(35),\, 440(40) \\
SAID(SE) 
&939(1),\, 0&1512(2),\, 113(6) & 1678(4),\, 113(3) & 2132(24),\, 550(25) & 2173(7),\, 445(21) \\
SAID(ED) 
&939(1),\, 0&1515(2),\, 109(4) & 1674(3),\, 114(7) & 2060(11),\, 521(16) & 2177(4),\, 464(9) \\
KH80 
&939(1),\, 0&1506(2),\, 115(3) & 1674(3),\, 129(4) & --- &2127(27),\, 380(29)\\
KA84 
&939(1),\, 0&1506(2),\, 116(4) & 1672(3),\, 132(5) & --- &2139(6),\, 390(7) \\
\end{tabular}
\end{ruledtabular}
\end{table*}
\begin{table*}
\caption{Summary of pole positions $M_p,\Gamma_p$ in MeV
for $I^\eta=\frac{1}{2}^-$ states. 
Notation as in Table~\ref{tab:regge12p}.}
\label{tab:regge12m}
\begin{ruledtabular}
\begin{tabular}{c|cccc}
Name &$N(1720)$ & $N(1675)$& $N(1990)$&
$N(2250)$\\
Status &**** & **** & ** & ****\\
$I^\eta_{(\tau)}\, J^P_p$& 
$\frac{1}{2}^-_{(-)}\, 3/2^+$ &
$\frac{1}{2}^-_{(+)}\, 5/2^-$ &
$\frac{1}{2}^-_{(-)}\, 7/2^+$ &
$\frac{1}{2}^-_{(+)}\, 9/2^-$ \\
\hline
CMB    
& 1680(30), \, 120(40) & 1660(10),\, 140(10) & 1900(30),\, 260(60)\phantom{0}&2150(50),\, 360(100) \\
J\"uBo 
& 1689(4),\, 191(3) & 1647(8),\, 135(9) & 2152(12), \, 225(20)\phantom{0} & 1910(53),\, 243(73) \\
BnGa 
& 1670(25),\, 430(100) & 1655(4),\, 147(5) & 1970(20),\, 250(20)\phantom{0} & 2195(45),\, 470(50) \\
SAID(SE) 
& 1668(24),\, 303(58) & 1661(1),\, 147(2.4) & 2157(62),\, 261(104) & 2283(10),\, 304(31) \\
SAID(ED) 
& 1659(11),\, 303(19) & 1657(3),\, 139(5)& --- & 2224(5),\, 417(10)  \\
KH80 
& 1677(5),\, 184(9) & 1654(2),\, 125(4) & 2079(13),\, 509(23) & 2157(17),\, 412(51) \\
KA84 
& 1685(5),\, 178(9) & 1656(1),\,123(3) & 2065(14),\, 526(9)\phantom{0} & 2187(7),\, 396(25) \\
\end{tabular}
\end{ruledtabular}
\end{table*}
\begin{table*}
\caption{Summary of pole positions $M_p,\Gamma_p$ in MeV
for $I^\eta=\frac{3}{2}^+$ states. 
Notation as in Table~\ref{tab:regge12p}.}
\label{tab:regge32p}
\begin{ruledtabular}
\begin{tabular}{c|cccc}
Name 
& $\Delta(1700)$ & $\Delta(1905)$ & $\Delta(2200)$& $\Delta(2300)$ \\
Status & **** & **** & *** & ** \\
$I^\eta_{(\tau)}\, J^P_p$& 
$\frac{3}{2}^+_{(-)}\, 3/2^-$ &
$\frac{3}{2}^+_{(+)}\, 5/2^+$ &
$\frac{3}{2}^+_{(-)}\, 7/2^-$ &
$\frac{3}{2}^+_{(+)}\, 9/2^+$ \\
\hline
CMB   
& 1675(25),\, 220(40) & 1830(40),\, 280(60) & 2100(50),\, 340(80) & 2370(80),\, 420(160) \\
J\"uBo 
& 1667(28),\, 305(45)& 1733(47),\, 435(264) & 2290(132),\, 388(204) & ---\\
BnGa 
& 1685(10),\, 300(15) & 1800(6),\, 290(15) & --- & --- \\
SAID(SE) 
& 1646(11),\, 203(17) & 1831(7),\, 329(17) & --- & --- \\
SAID(ED) 
& 1652(10),\, 248(28) & 1814(5),\, 273(9) & --- & --- \\
KH80 
& 1643(9),\, 217(18) & 1752(5),\, 346(8) & --- & --- \\
KA84 
& 1616(5),\, 280(9) & 1790(5),\, 293(12) & --- & --- \\
\end{tabular}
\end{ruledtabular}
\end{table*}
\begin{table*}
\caption{Summary of pole positions $M_p,\Gamma_p$ in MeV
for $I^\eta=\frac{3}{2}^-$ states. 
Notation as in Table~\ref{tab:regge12p}.}
\label{tab:regge32m}
\begin{ruledtabular}
\begin{tabular}{c|ccccc}
Name & $\Delta(1232)$ & $\Delta(1930)$ & $\Delta(1950)$& ---& $\Delta(2420)$\\
Status & **** & *** & **** & --- & ****\\
$I^\eta_{(\tau)}\, J^P_p$& 
$\frac{3}{2}^-_{(-)}\, 3/2^+$ &
$\frac{3}{2}^-_{(+)}\, 5/2^-$ &
$\frac{3}{2}^-_{(-)}\, 7/2^+$ &
$\frac{3}{2}^-_{(+)}\, 9/2^-$ &
$\frac{3}{2}^-_{(-)}\, 11/2^+$ \\
\hline
CMB    
&1210(1),\, 100(2)&1890(50),\, 260(60)&1890(15),\, 260(40) &--- & 2360(100),\, 420(100)\\
J\"uBo 
&1215(4),\, 97(2)& 1663(43),\, 263(76)&1850(37),\, 259(61) & 1783(86),\, 244(194)& --- \\
BnGa 
&1210.5(1.0),\, 99(2)&---&1888(4),\, 245(8) & ---& ---\\
SAID(SE) 
&1211(0),\, 100(2) & 1845(31),\, 174(40) & 1888(3),\, 234(6)& --- & --- \\
SAID(ED) 
&1211(2),\, 98(3)&1969(23),\, 248(36)&1878(4),\, 227(6)&1955(24),\, 911(24)&2320(13),\, 442(23) \\
KH80 
&1211(2),\, 98(3) & 1848(28),\, 321(24) & 1877(3),\, 223(5) & --- & 2454(15),\, 462(58) \\
KA84 
&1210(2),\, 100(2) & 1844(36),\, 334(26) & 1878(3),\, 246(7) & --- & 2301(7),\, 533(17) \\
\end{tabular}
\end{ruledtabular}
\end{table*}

For a given spin and parity, 
resonance pole positions 
$s_p$ are extracted
from partial wave amplitudes analytically continued 
off the real energy axis to the unphysical Riemann sheet. 
On the real axis the partial wave amplitudes are fitted 
to the data on meson-nucleon scattering 
and meson photoproduction. 
This procedure carries uncertainties associated to the
experimental data (systematic and statistical),
the partial wave analysis model itself, 
and the analytic continuation to the
complex energy plane.
The differences among models in the pole extractions 
reflect on some of these uncertainties and model dependencies.
In Tables~\ref{tab:regge12p}-\ref{tab:regge32m} we list the
poles that, in principle,
conform the leading (parent),
{\it i.e.} the trajectory composed by
the lowest mass states
for each spin-parity assignment,
$N^*$ and $\Delta^*$ Regge trajectories,
classified according to 
isospin $I$, 
naturality $\eta$
($\eta=+1$ if $P=(-1)^{J_p-1/2}$ 
and $\eta=-1$ if $P=-(-1)^{J_p-1/2}$
where $P$ is the parity and $J_p$ is the spin of the resonance),
and signature $\tau$ ($\eta=\tau P$).
The quantum numbers 
identify  a given $I^\eta_{(\tau)}$ trajectory, 
{\it e.g.} the trajectory which contains 
$N(939)$ (the nucleon) corresponds to  
$I^\eta_{(\tau)}=\frac{1}{2}^+_{(+)}$.
We note that out of the four trajectories, three do not
contain the lowest spin $1/2$ resonance. 
States are absent for dynamical reasons. 
For example, in the case of the
$I^\eta=\frac{3}{2}^-$ trajectory it is unlikely that QCD yields   
a spin $1/2^-$ state with lower mass than the $\Delta(1232)$. Therefore  
 the isospin $3/2$ spin $1/2^-$,  $\Delta(1620)$, 
 has to be associated with a daughter trajectory. 
In the $\frac{1}{2}^-$ parent trajectory, the 
four-star $N(1535)$ $1/2^-$ resonance
could be a candidate for the lowest spin state,  however, its position on the Chew-Frautschi plot, where it aligns
with the  $N(1900)$ $3/2^+$ and $N(2060)$ $5/2^-$ states, 
 makes it a better fit
 with the first daughter trajectory.  
Finally, the one-star $\Delta(1750)$ $1/2^+$ and
the four-star $\Delta(1910)$ $1/2^+$ are most  likely on a daughter, since their  masses are 
 higher than $\Delta(1700)$ $3/2^-$
, which appears on the parent trajectory. 
Phenomenologically, it is observed that the leading 
Regge trajectories that differ only by
signature are (almost) degenerate,
{\it i.e.}
odd ($\tau=-$) and even ($\tau=+$) signatures
have the same trajectory. 
For subleading trajectories there is often not 
enough information to disentangle both signatures. 
We use seven sets of resonance poles 
extracted from the following analyses:
\begin{itemize}
\item[(i)] \textbf{CMB}: 
Pole parameters from the Carnegie-Mellon-Berkeley 
$\pi N$ partial wave analysis 
of~\cite{Cutkosky:1979fy,Cutkosky:1980rh}
as quoted by the PDG~\cite{Tanabashi:2018oca};
\item[(ii)] \textbf{J\"uBo}: 
Pole parameters from ~\cite{Ronchen:2018ury} using the
J\"ulich-Bonn 2017 coupled-channel model.  
The resonance spectrum is obtained from a combined analysis 
of $\eta$, $\pi$ and $K\Lambda$ 
photoproduction off 
the proton together with the reactions 
$\pi N$ $\to \pi N$, $\eta N$, $K \Lambda$ 
and $K\Sigma$;
\item[(iii)] \textbf{BnGa}: 
Pole parameters given
in~\cite{Anisovich:2011fc,Sokhoyan:2015fra}
from the Bonn-Gatchina multichannel partial wave analysis 
of $\pi N$ elastic scattering 
data and pion and photo-induced inelastic 
reactions;
\item[(iv)] \textbf{SAID(SE)}: 
Pole parameters obtained in~\cite{Svarc:2014aga}
from a fit to the single-energy SAID-GW WI08 partial waves of
$\pi N$ elastic scattering~\cite{Workman:2012hx}
using the Laurent$+$Pietarinen (LP) approach;
\item[(v)] \textbf{SAID(ED)}: 
Poles extracted in~\cite{Svarc:2014aga} 
from the energy-dependent 
SAID-GW WI08 partial waves of
$\pi N$ elastic scattering~\cite{Workman:2012hx}
also using the LP approach;
\item[(vi)] \textbf{KH80}: 
Pole extracted in~\cite{Svarc:2014zja}
from the Karlsruhe-Helsinki 
KH80~\cite{LandoltBornstein1983:sm_lbs_978-3-540-39059-6_90}
partial wave analysis of $\pi N$ 
elastic scattering employing 
the LP approach; and
\item[(vii)] \textbf{KA84}: 
Pole extracted in \cite{Svarc:2014zja} 
from the Karlsruhe KA84~\cite{Koch:1985bp,Koch:1985bn}
partial wave analysis of $\pi N$ elastic scattering
employing the LP approach.
\end{itemize}
Other pole extractions are available in the literature. These include, 
the speed plot extraction from $\pi N\to \pi N$
amplitudes by H\"ohler~\cite{Hoehler1993};
the SAID pole parameters given 
in~\cite{Svarc:2014aga} obtained from
the SAID-GW WI08 partial wave analysis 
of $\pi N$ elastic scattering~\cite{Workman:2012hx};
the Kent State University (KSU) 
pole extraction in~\cite{Shrestha:2012ep}
using a multichannel parametrization of 
$\pi N$ scattering amplitudes;
the Pittsburgh-Argonne National Lab (P-ANL) 
pole extraction in~\cite{Vrana:1999nt};
the Giessen group 
coupled-channel analysis of $\eta$ 
production
and photoproduction data 
on the proton~\cite{Shklyar:2012js};
the Argonne National Lab-Osaka (ANL-O)
amplitude analysis of
$\pi N\to \pi N$, $\eta N$, $K\Lambda$,
$K\Sigma$ and $\gamma N \to$ $\pi N$, 
$\eta N$,$K\Lambda$, $K\Sigma$ data~\cite{Kamano:2013iva};
and the Zagreb analysis in~\cite{Batinic:2010zz} 
based on the CMB coupled-channel approach;
H\"ohler, SAID, KSU, P-ANL, Giessen and ANL-O 
do not provide uncertainties
in their pole extractions and the Zagreb group analysis only
studies the $N^*$ spectrum,
hence, we choose not to include them in our work.
Also, we do not include superseded pole extractions 
within the same reaction models.

\begin{figure}
\centering
\subfigure[\ $N^*$ resonances.]{
\rotatebox{0}{\scalebox{0.31}[0.31]{\includegraphics{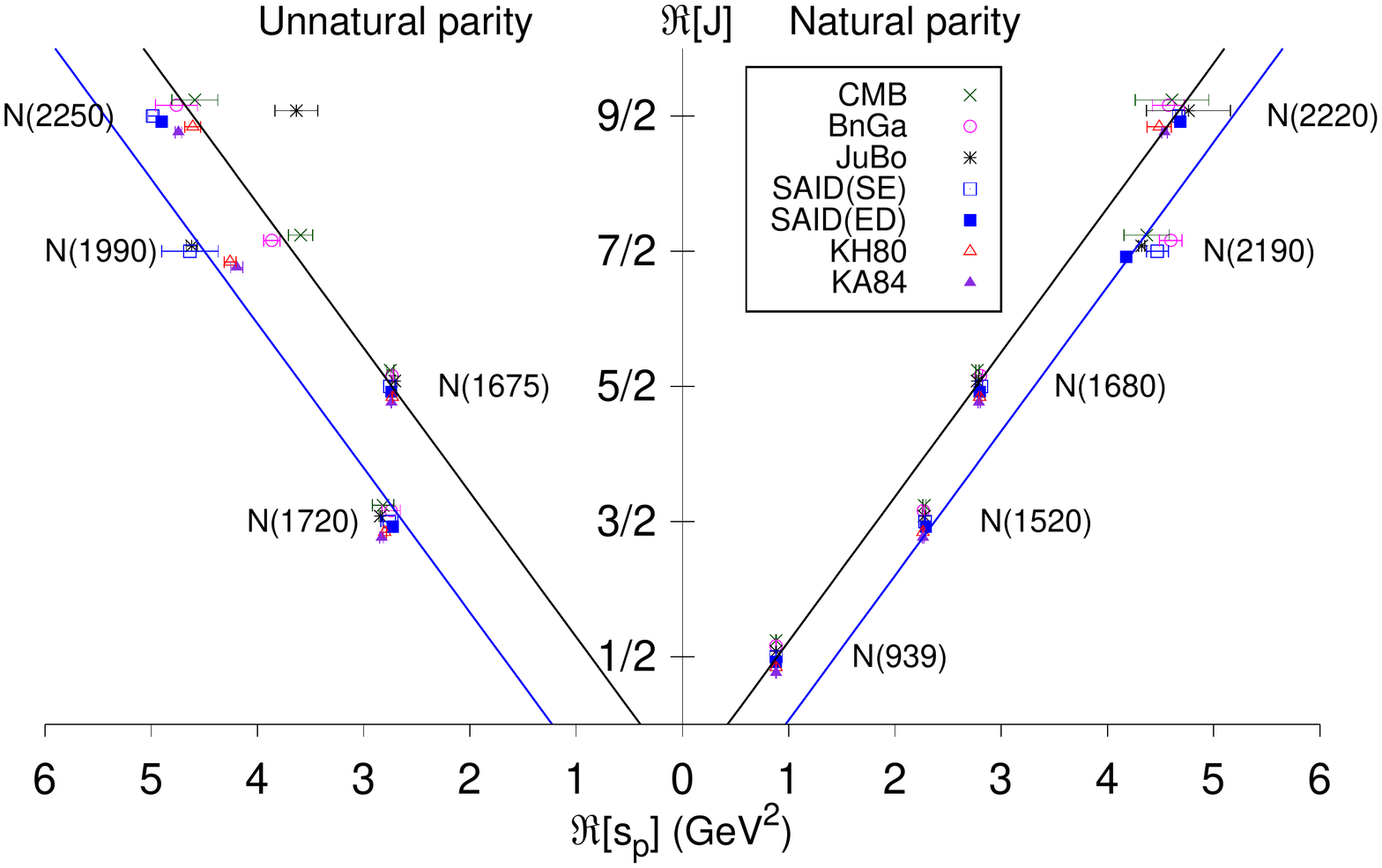}}}
\label{fig:poles12rega}} 
\subfigure[\ $\Delta^*$ resonances.]{
\rotatebox{0}{\scalebox{0.31}[0.31]{\includegraphics{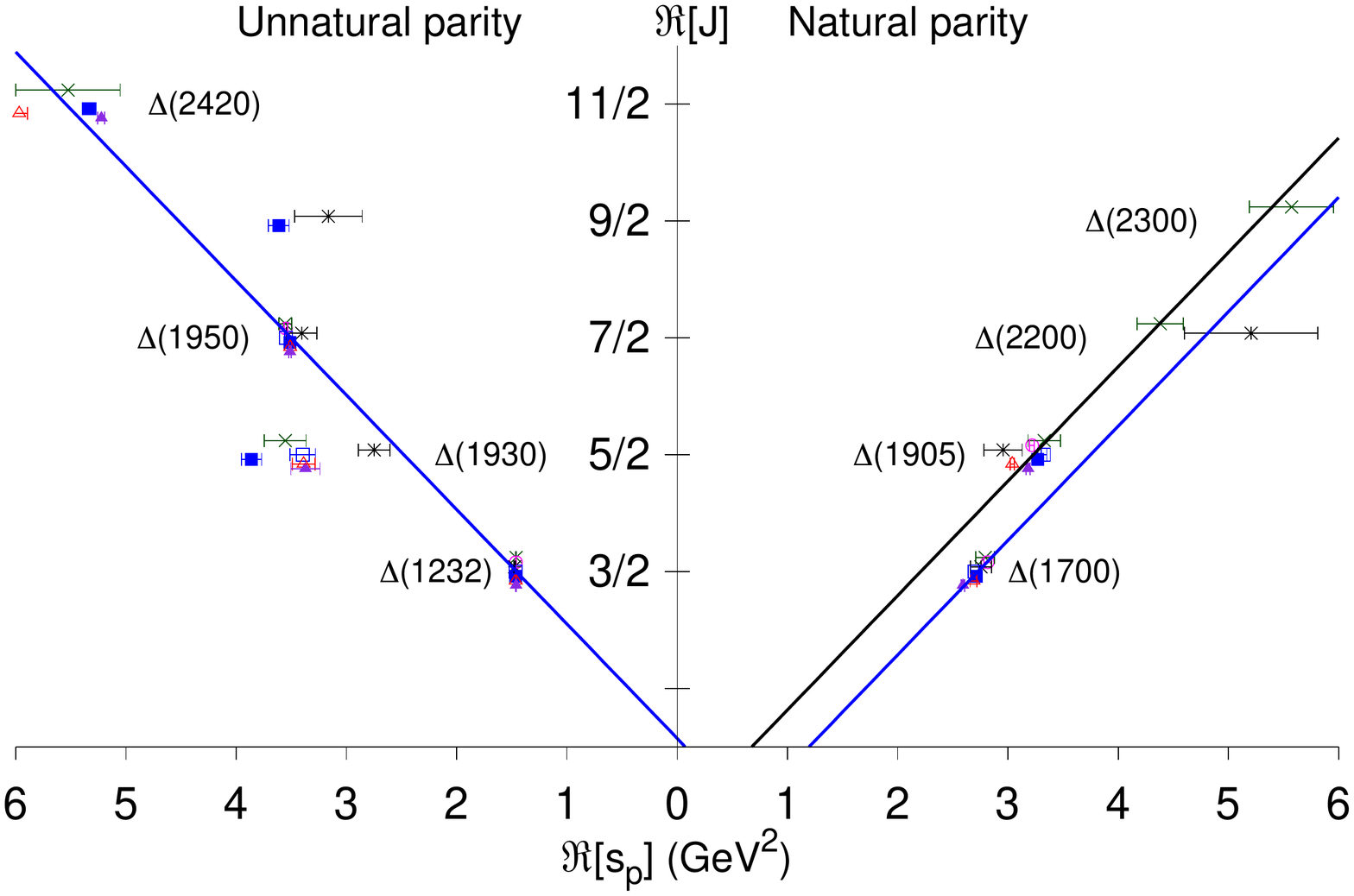}}}
\label{fig:poles32rega}} 
\caption{Chew--Frautschi plots 
for the leading $N^*$  and $\Delta^*$ Regge trajectories in 
Tables~\ref{tab:regge12p}-\ref{tab:regge32m}.
Solid black (blue) curves
are linear fits to the displayed positive (negative) signature data points
(see Sec.~\ref{sec:poles} for details.)
All the curves share the same slope as required by
MacDowell symmetry~\cite{MacDowell:1959zza}.
We do not show a fit for the $\frac{3}{2}^+_{(+)}$ states 
because the $\Delta\: 9/2^-$ pole is
unreliable as will be discussed in Sec.~\ref{sec:32mregge_trajectory}.
In order to make the plots readable, 
the poles are slightly displaced from the
correct $\Re[J]=J_p$ value.
} 
\label{fig:polesrega}
\end{figure}
\begin{figure}
\centering
\subfigure[\ $N^*$ resonances.]{
\rotatebox{0}{\scalebox{0.31}[0.31]{\includegraphics{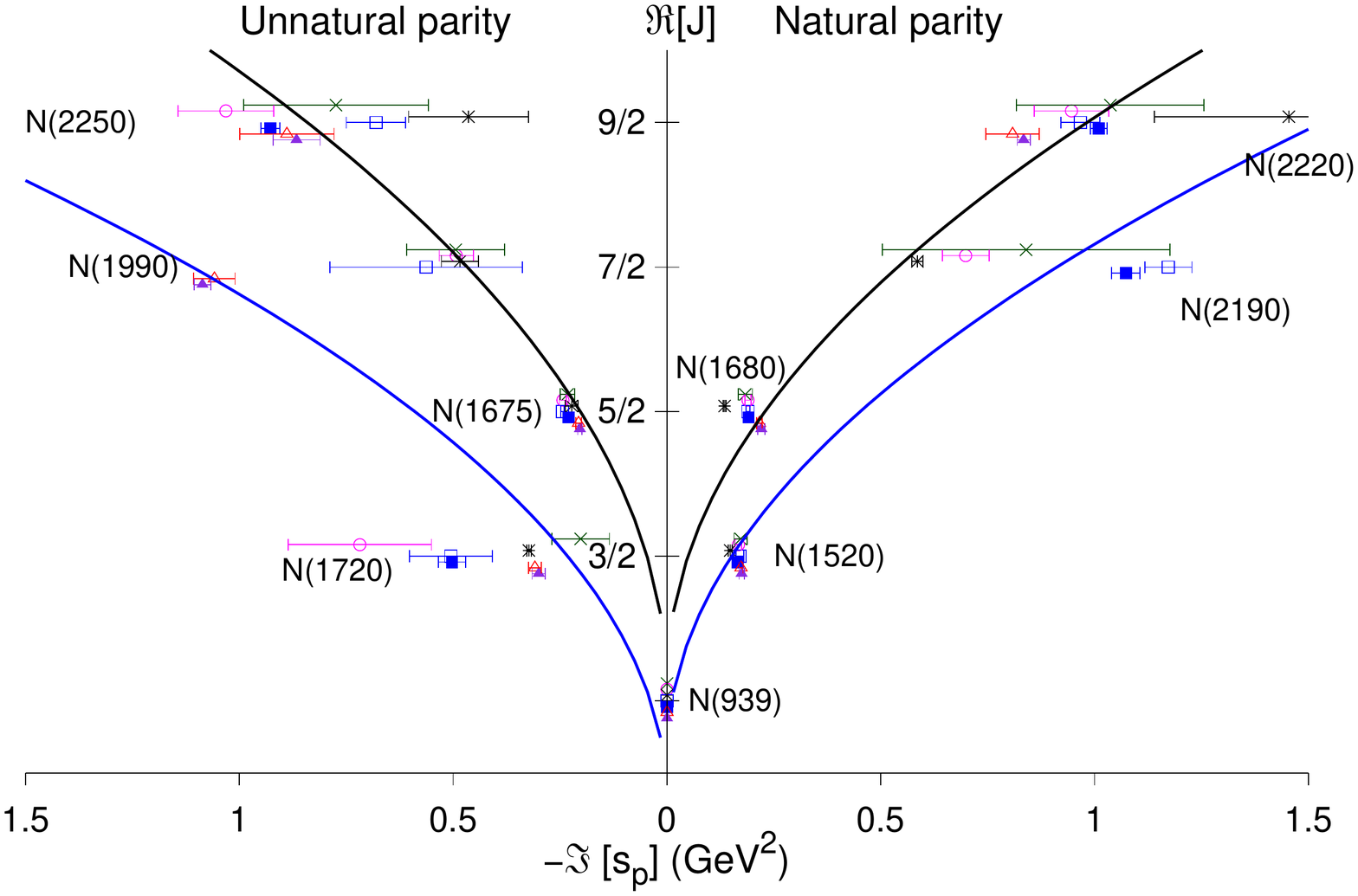}}}
\label{fig:poles12regb}} 
\subfigure[\ $\Delta^*$ resonances.]{
\rotatebox{0}{\scalebox{0.31}[0.31]{\includegraphics{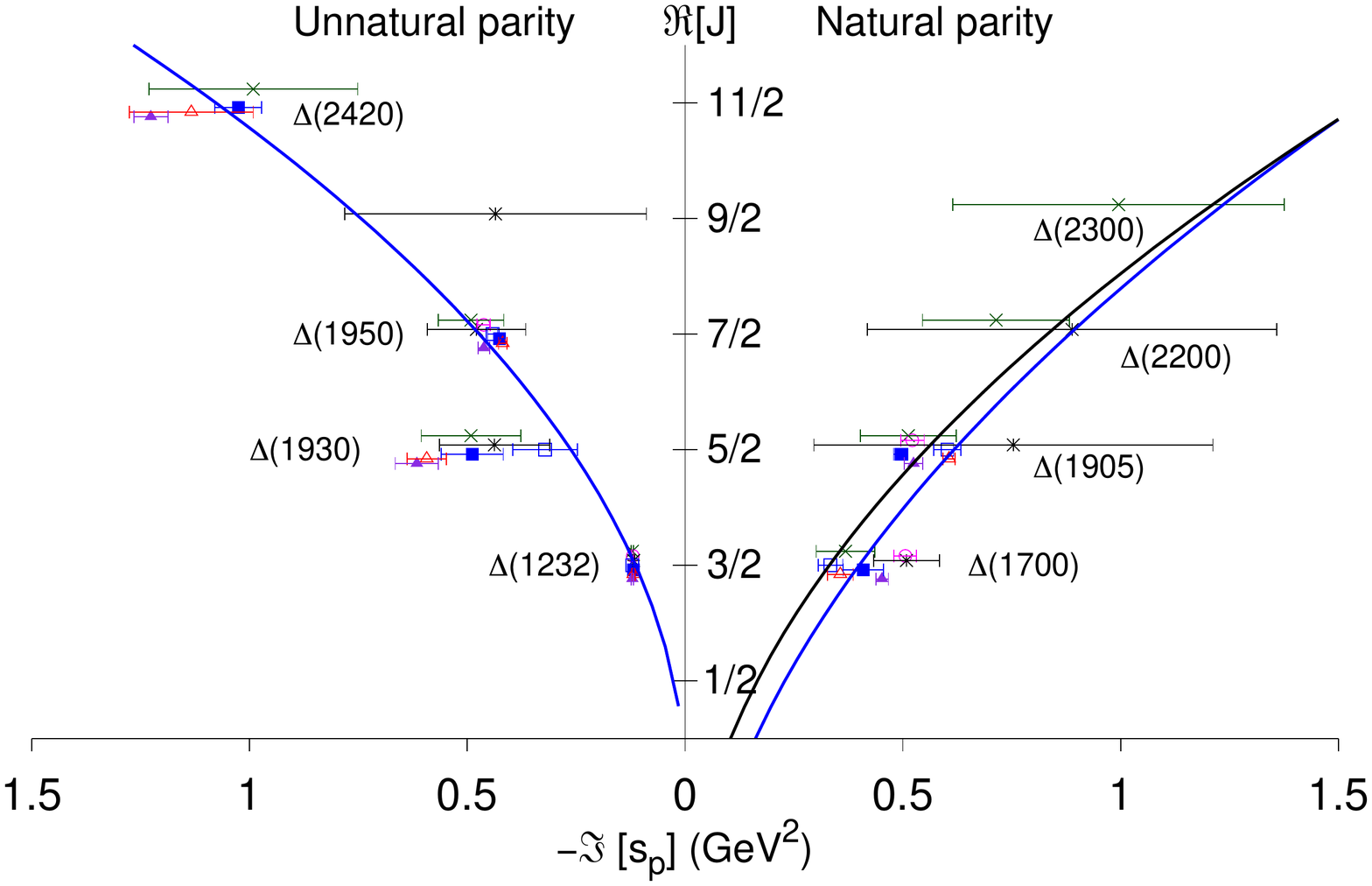}}}
\label{fig:poles32regb}} 
\caption{$(\Im[s_p],\Re[J]=J_p)$ plots 
introduced in~\cite{Fernandez-Ramirez:2015fbq}
for the leading $N^*$ and $\Delta^*$ Regge trajectories in 
Tables~\ref{tab:regge12p}-\ref{tab:regge32m}.
Solid black (blue) curves
are square-root  fits to the displayed positive (negative) signature data points 
(see Sec.~\ref{sec:poles}  for details.)
We do not show a fit for the $\frac{3}{2}^+_{(+)}$ states 
because the $\Delta\: 9/2^-$ pole is
unreliable as will be discussed in Sec.~\ref{sec:32mregge_trajectory}.
The different pole sets 
are labeled as in Fig.~\ref{fig:polesrega}. 
In order to make the plots readable, 
the poles are slightly displaced from the
correct $\Re[J]=J_p$ value as in Fig.~\ref{fig:polesrega}.
SAID(ED) $\Delta$ $9/2^-$ pole 
in the unnatural parity trajectory
has a very large $\Im[s_p]$ value and it is not shown
in plot (b).
} 
\label{fig:polesregb}
\end{figure}

In Fig.~\ref{fig:polesrega}  
we show the Chew-Frautschi plots 
$(\Re[s_p],\Re[J]=J_p)$
for the $N^*$ and $\Delta^*$ resonances,
and Fig.~\ref{fig:polesregb}
displays the
$(\Im[s_p],\Re[J]=J_p)$ plots 
introduced in~\cite{Fernandez-Ramirez:2015fbq}.
These figures provide a qualitative description 
of the spectrum. We note the spectrum exhibits the 
approximate linear behavior in  
$(\Re[s_p],J_p)$
and the square root-like behavior in 
$(\Im[s_p],J_p)$. This was also observed  
in the spectrum of 
the hyperons~\cite{Fernandez-Ramirez:2015fbq}.
To highlight the linear 
trend of the poles in Fig.~\ref{fig:polesrega} 
we show linear fits, $J_p=a+b\: \Re[s_p]$, to
each $\frac{1}{2}^\pm_{(\pm)}$, $\frac{3}{2}^\pm_{(-)}$, and $\frac{3}{2}^-_{(+)}$
Chew-Frautschi plot with a common slope $b$
as required by MacDowell symmetry~\cite{MacDowell:1959zza}.
We do not show a fit for the $\frac{3}{2}^+_{(+)}$ states 
because the $\Delta\: 9/2^-$ pole is
unreliable as will be discussed in Sec.~\ref{sec:32mregge_trajectory}.
To highlight the square-root trend of the poles in the $(\Im[s_p],J_p)$ plots
we show square-root fits, $J_p=c+d\: \sqrt{-\Im[s_p]}$,
to each $\frac{1}{2}^\pm_{(\pm)}$, $\frac{3}{2}^\pm_{(-)}$, and $\frac{3}{2}^-_{(+)}$
set of states in Fig.~\ref{fig:polesregb}
with $c$ and $d$ parameters unconstrained.
We remark that these fits to $(\Re[s_p],J_p)$ and 
$(\Im[s_p],J_p)$ plots were performed separately.
Hence, we do not provide further information on these naive fits as they
are just exploratory computations to remark the linear and square-root trends of the poles
in the plots.
A quantitative analysis of the Regge trajectories has to be performed
within a model fitting to the real and imaginary parts of the poles
simultaneously as it is done in the analysis that follows.
We defer  the rest of the discussion of 
 these plots to Sec.~\ref{sec:results}, 
where we present the quantitative analysis of the spectrum.

\section{Models for the parent Regge trajectories}
\label{sec:models}
In what follows the working hypothesis is that
the square-root-like behavior displayed in
Fig.~\ref{fig:polesregb}
is the leading singularity of the trajectories 
as implied by unitarity~\cite{Chew:1960iv}. 
This stems from the fact that the leading two-body decay
channels,{\it i.e.} those
that account for most of the cross section,
give the imaginary part proportional to the relative 
momentum $q\sim\sqrt{s-s_t}$,
where $s$ is the two-body invariant mass squared 
and $s_t$ is the threshold. Contribution from multi-body  
final states can effectively be absorbed into model 
parameters. 
Near a Regge pole, partial wave amplitudes 
are proportional to 
\begin{equation}
t_\ell(s) \propto \frac{1}{\ell-\alpha(s)},
\label{eq:poleamplitude}
\end{equation}
where $\alpha (s)$ is the Regge trajectory 
and $\ell$ 
is the total angular momentum of the 
partial wave, that matches the spin 
$J_p$ of the resonance. 
This can be compared  to the 
Breit-Wigner amplitude close to the $s_p$ pole
under the approximation of elastic two-body 
scattering,\footnote{We note that both
Eqs.~\eqref{eq:poleamplitude}
and~\eqref{eq:bw} are written in the 
second Riemann sheet of the 
complex $s$ plane, where the resonant poles 
in the amplitude appear.} 
\begin{equation}
t_\ell(s) \propto \frac{g^2}{M^2-s-i \, g^2\rho(s,s_t)},
\label{eq:bw}
\end{equation}
where $M$ is real, sometimes referred to as the 
Breit-Wigner mass. 
Resonance decay is determined by $g^2$, which can be used to  
define coupling to open channels and  $\rho(s,s_t)$ 
which is the phase space factor. With the 
determination of $\rho(s,s_t)$ that is analytical across the real axis 
for $s > s_t$ one finds 
poles of $t_\ell(s)$ located on the lower half $s$-plane 
that are analytically connected to the physical region at $s + i\epsilon$. 
How deep a pole is in the complex plane depends on two
factors, the dynamics of QCD and the phase space.
The phase space dependence $\rho(s,s_t)$ is 
explicitly built in through unitarity and QCD dynamics
are hidden in the parameters, $M$ and $g$.
At the pole  $s_p$, 
Eqs.~\eqref{eq:poleamplitude} and~\eqref{eq:bw} 
have to be equal, hence
\begin{equation}
 \ell-\alpha(s_p)=
 \frac{M^2}{g^2}-\frac{s_p}{g^2}-i\rho(s_p,s_t)=0\,.
\label{eq:poleequality}
\end{equation}
This equation is used to relate the 
imaginary part of the Regge trajectory 
to resonance decay parameters.
Without loss of generality, we can parametrize 
the Regge trajectory 
as~\cite{Chiu:1972xu,Mandelstam:1968zza,Fernandez-Ramirez:2015fbq}
\begin{equation}
\alpha(s) = \alpha_0 + \alpha' s 
+ i\, \gamma \, \phi(s,s_t) \, ,
\label{eq:regge}
\end{equation}
where $\alpha_0$, $\alpha'$ and $\gamma$ are real constants,
and $\phi(s,s_t)$ contains information about resonance decay.
The slope $\alpha'$ is often 
related to the tension of the confining string in flux tube
models~\cite{Isgur:1984bm,Olsson:1995mv,Semay:2007cv}
and to the range of the strong interaction in Veneziano models~\cite{Veneziano:1968yb}.
The square-root-like behavior in Fig.~\ref{fig:polesregb}
hints that $\rho(s,s_t)$ is the dominant component
of $\phi(s,s_t)$.
As previously stated, 
the position of the pole in the complex plane depends on
the dynamics of QCD and the phase space,
so, both contribute to the functional form of $\phi(s,s_t)$.
As a first approximation, 
we can model $\gamma \phi(s,s_t)=\rho(s,s_t)$,
and fit the trajectory in Eq.~\eqref{eq:regge} at
the poles $s=s_p$ to
$\Re \left[ \alpha(s_p) \right]= \Re [J]=J_p$ 
and $\Im \left[ \alpha(s_p) \right]=\Im[J]=\Im[J_p]= 0$
obtaining $\alpha_0$, $\alpha'$, $\gamma$ 
and $s_t$.
The parameter $\alpha_0$ is dimensionless, 
the slope $\alpha'$ has units of GeV$^{-2}$, 
$s_t$ acts as an effective threshold 
that has units of GeV$^2$.
In this way, $\phi(s,s_t)$ has the phase space
contribution to the pole position explicitly build in, 
and any difference with the actual functional form of the Regge trajectory
has to be due to additional QCD dynamics.
The systematic uncertainties of the model associated with the
description of the phase space factor far away 
from the threshold 
can be studied by considering different models for 
$\phi(s,s_t)$. In particular we use, 
\begin{subequations}
\begin{alignat}{2}
i\, \phi_{0}(s,s_t) = &\, 0 \, , 
\label{eq:models0}\\
i\, \phi_\text{I}(s,s_t) = &\,  i\, \sqrt{s-s_t} \, ,
\label{eq:modelsI}\\
i\, \phi_\text{II}(s,s_t) =&\,
i\beta(s,s_t) + 2 i \tau(s,s_t),
\label{eq:modelsIII}
\end{alignat}\label{eq:models}
\end{subequations}
where 
\begin{alignat}{2}
i\beta(s,s_t)=&\frac{s-s_t}{\pi}
\int_{s_t}^\infty \frac{\tau(s',s_t)}{s'-s_t}  
\frac{ds'}{s'-s} \nonumber \\
=&\frac{2}{\pi}\frac{s-s_t}{\sqrt{s(s_t-s)}}
\arctan \sqrt{\frac{s}{s_t-s}} \, ,
\end{alignat}
is the analytic continuation of the two-body 
phase space\footnote{
We assume elastic two-body scattering, 
and hence, all poles are considered 
to be in the second Riemann sheet.
That is also the reason why 
we fit an effective threshold $s_t$ 
instead of using the actual physical thresholds.}
$\tau(s,s_t)=\sqrt{1-s_t/s}$
to the complex $s$ plane. It follows that in Eq.~\eqref{eq:regge}, 
$\gamma$  has units of GeV$^{-1}$ 
for model I and is dimensionless 
in model II.
Model 0 is the customary linear dependency that ignores
the existence of the imaginary part
of the resonance poles.
Although essential physics is ignored in such model, 
we fit it to $\Re \left[ s_p\right]$ for completeness and 
to provide a comparison to previous works.
We note that once the width of the resonance pole 
is taken into account
it is clear that a Regge trajectory cannot be linear. 
Linear Regge trajectories can only happen for zero-width
resonances, {\it e.g.} resonances computed as bound states
in a constituent quark model, or the tower of states in 
the Veneziano amplitude~\cite{Szczepaniak:2014bsa}.
Models I and II do incorporate such physics
by adding an imaginary part to $\alpha(s)$ in a simple way.
Model I is a customary approach to
add the imaginary part to $\alpha(s)$
which has been used to
account for unitarity effects in Veneziano-type
amplitudes~\cite{Bugrij:1973ph,Jenkovszky:1974hu,Shi:2014nea}.
Model II is the most physically motivated as it is 
guided
by the relation between Eqs.~\eqref{eq:poleamplitude}
and~\eqref{eq:bw}, $\beta(s,s_t)$
is the analytic continuation of the phase space,
Chew-Mandelstam dispersive approach~\cite{Chew:1960iv},
and $\phi(s,s_t)$ is the analytic continuation of 
$\beta(s,s_t)$ to the second Riemann sheet, 
as dictated by unitarity.
However, we will compute the three models 
for the sake of completeness
and comparison purposes.

Our hypothesis to interpret the nature of the
resonances in terms of the
Regge trajectory is that
a state that is located on a linear 
trajectory in the 
Chew-Frautschi plot {\it and} a square-root-like behavior
in $(\Im \left[ s_p\right],J_p)$ plot
would be mostly a compact $3q$ state candidate. 
Hence, most of the width, {\it i.e.} the contribution to
$\phi(s,s_t)$, would be due to the phase space.
This assessment  can by strengthen  by a more 
 quantitative analysis in which we fit the poles to the models in Eq.~\eqref{eq:models}.
If the states are truly $3q$ states,
the poles should adhere nicely to our
Regge trajectory models, 
{\it i.e.} phase space dominates how deep the pole is in the complex plane and there is little room for additional
QCD dynamics.
If the resonance pole is not well described by
our models, 
it is an indication that additional QCD
dynamics are important, signaling that the state
has significant physics beyond the compact $3q$ picture.
To summarize, the way we proceed in the quantitative analysis is as follows:
(i) We fit the poles in a given trajectory to the models; 
(ii) on average the description must be approximately correct because of the linear and square-root-like behaviors;
(iii) however, our Regge trajectory model only accounts for the {\it phase space} contribution, so it is incomplete;
(iv) deviations from the models are associated to the physics that our model lacks, {\it i.e.} additional QCD dynamics, 
which we interpret as physics that go beyond the $3q$ picture.

\section{Results}
\label{sec:results}
\subsection{Fits and error analysis}
\label{sec:fits}
To determine the parameters 
$\alpha_0$, $\alpha'$, $\gamma$ and $s_t$ in 
Eq.~\eqref{eq:regge} for a given pole extraction
we use the least-squares method 
by minimizing the distance squared $d^2$ 
between the trajectory $\alpha(s)$ evaluated at the complex 
pole position $s_p$ and the angular momenta $J$,
\begin{equation} 
d^2 = \sum_{poles} \{ \,
\left[\, \Re[J]- \Re [\alpha(s_p)] \, \right]^2 
+  \left[ \Im[J]- \Im [ \alpha(s_p)]\,  \right]^2 \, \}\, .
\nonumber
\end{equation}
with $\Re[J] = J_p$ and 
$\Im[J] = \Im[J_p]=0$ for the resonance poles. 
The value of $s_t$ should be 
compatible with its interpretation as an
effective threshold in the resonance region.
This is used as the criterion to select the
physically meaningful minimum if several 
local minima appear in the fits.
We estimate the errors in the parameters
through the bootstrap technique~\cite{recipes,EfroTibs93,Landay:2016cjw}.
In doing so, we perform $10^4$ fits to 
pseudodata generated according to
the pole uncertainties.
The expected value of each parameter 
is computed as the mean of the 
$10^4$ samples and the uncertainty 
is given by the standard deviation. 
This method is described in detail 
in~\cite{Fernandez-Ramirez:2015fbq,Molina:2017iaa}
and allows to propagate the uncertainties 
from the poles to the parameters accounting
for all the correlations.
The systematic errors associated 
with model dependence in the amplitude analyses are not  considered in the pole extractions, hence, 
we take the differences among models as an indication
of such uncertainties.
The fit results are provided and discussed in
Sec.~\ref{sec:regge_trajectory}.

\subsection{Regge trajectories}
\label{sec:regge_trajectory}
\subsubsection{$\frac{1}{2}^+$ Regge trajectory}
\label{sec:12pregge_trajectory}
\begin{table}
\caption{Parameter $\alpha_0$ obtained for $\frac{1}{2}^+$ 
trajectories and models 0, I and II.}
\label{tab:nat12palpha0}
\begin{ruledtabular}
\begin{tabular}{clccc}
 $I^\eta_{(\tau)}$ &Pole set
 &$\alpha_0^{(0)}$ &$\alpha_0^{(\text{I})}$  & $\alpha_0^{(\text{II})}$\\
\hline
 $\frac{1}{2}^+_{(+)}$   
 & CMB& 
$-0.4(1)\phantom{0}$& $\phantom{-}0.3(2)\phantom{0}$& $\phantom{-}0.3(3)\phantom{0}$\\
& J\"uBo& 
$-0.3(1)\phantom{0}$& $\phantom{-}0.6(1)\phantom{0}$& $\phantom{-}0.9(3)\phantom{0}$\\
 & BnGa     &
 $-0.46(5)$& $\phantom{-}0.20(7)$& $\phantom{-}0.1(2)\phantom{0}$\\
 & SAID(SE) & $-0.42(1)$& $\phantom{-}0.25(3)$ &$\phantom{-}0.22(6)$\\
 & SAID(ED) & $-0.41(1)$& $\phantom{-}0.29(2)$ &$\phantom{-}0.30(3)$\\
 & KH80     & $-0.50(4)$& $-0.1(2)\phantom{0}$& $-0.2(1)\phantom{0}$\\
 & KA84     & $-0.48(1)$& $\phantom{-}0.05(3)$& $-0.09(3)$\\
 \hline
 $\frac{1}{2}^+_{(-)}$   
 & CMB    & $-0.6(1)\phantom{0}$& $-0.8(3)\phantom{0}$& $-3.5(7)\phantom{0}$\\
 & J\"uBo   & $-0.71(3)$& $-0.79(4)$& $-1.53(6)$\\
 & BnGa     & $-0.44(7)$& $-0.53(7)$& $-1.5(5)\phantom{0}$\\
 & SAID(SE) & $-0.53(7)$& $-0.9(1)\phantom{0}$ &$-4.6(3)\phantom{0}$\\
 & SAID(ED) & $-0.86(4)$& $-1.25(6)$&$-5.54(3)$\\
\end{tabular}
\end{ruledtabular}
\end{table}
\begin{table}
\caption{Parameter $\alpha'$ 
obtained for $\frac{1}{2}^+$ trajectories.} 
\label{tab:nat12palphaprime}
\begin{ruledtabular}
\begin{tabular}{clccc}
$I^\eta_{(\tau)}$ & Pole set
&$\alpha'^{(0)}$ &$\alpha'^{(\text{I})}$ & $\alpha'^{(\text{II})}$\\
\hline
$\frac{1}{2}^+_{(+)}$ 
& CMB &$1.06(7)$& $0.85(6)$&$0.9(1)\phantom{0}$\\
&J\"uBo & $1.00(8)$& $0.72(6)$&$0.8(1)\phantom{0}$\\
& BnGa &$1.07(3)$& $0.87(3)$&$1.04(6)$\\
&SAID(SE) & $1.04(1)$& $0.85(1)$&$0.99(1)$\\
& SAID(ED) &$1.036(4)$& $0.84(1)$&$0.97(1)$\\%
& KH80 &$1.10(2)$& $0.98(6)$&$1.14(5)$\\
& KA84 &$1.08(1)$& $0.93(1)$&$1.10(1)$\\ 
\hline
$\frac{1}{2}^+_{(-)}$ 
& CMB &$0.94(7)$& $0.95(9)$&$1.6(2)\phantom{0}$\\
& J\"uBo &$0.97(1)$& $0.98(1)$&$1.23(2)$\\
&BnGa &$0.85(3)$& $0.86(3)$&$1.15(6)$\\
& SAID(SE) & $0.89(3)$& $0.92(3)$&$2.0(1)\phantom{0}$\\
& SAID(ED) & $1.03(2)$& $1.06(2)$&$2.27(2)$\\
\end{tabular}
\end{ruledtabular}
\end{table}
\begin{table}
\caption{Parameters $\gamma$ and $s_t$
obtained for $\frac{1}{2}^+$ trajectories.}
\label{tab:nat12pg}
\begin{ruledtabular}
\begin{tabular}{clcccc}
 $I^\eta_{(\tau)}$ 
 & Pole set
 &$\gamma^{(\text{I})}$& $\gamma^{(\text{II})}$&$s_t^{(\text{I})}$ &$s_t^{(\text{II})}$\\
\hline
 $\frac{1}{2}^+_{(+)}$ 
 & CMB &$0.49(7)$&$0.66(7)$& $2.4(2)\phantom{0}$& $1.04(9)$\\
  &J\"uBo & $0.62(8)$& $0.67(5)$&$2.65(5)$&$1.3(1)\phantom{0}$\\
 & BnGa &$0.46(3)$&$0.65(4)$& $2.4(1)\phantom{0}$& $0.96(3)$\\
 &SAID(SE) & $0.46(2)$& $0.64(2)$&  $2.44(3)$& $0.98(1)$\\
& SAID(ED) &$0.48(1)$& $0.65(1)$& $2.46(3)$& $1.00(1)$ \\
 & KH80 &$0.39(3)$& $0.65(3)$& $1.8(4)\phantom{0}$& $0.91(1)$\\
 & KA84 &$0.41(1)$& $0.64(1)$& $2.06(7)$& $0.92(1)$ \\
 \hline
 $\frac{1}{2}^+_{(-)}$ 
 & CMB &$0.5(2)\phantom{0}$&$1.9(5)\phantom{0}$& $2.3(4)\phantom{0}$& $2.9(6)\phantom{0}$\\
 & J\"uBo &$0.39(1)$& $0.95(3)$& $2.17(2)$&$2.34(1)$\\
& BnGa &$0.38(3)$&$1.0(1)\phantom{0}$& $2.17(3)$& $2.42(4)$\\
& SAID(SE) & $0.72(5)$&$3.0(2)\phantom{0}$ & $2.39(2)$& $2.79(2)$\\
 & SAID(ED) & $0.82(3)$&$3.15(5)$& $2.40(1)$&$2.78(3)$\\
\end{tabular}
\end{ruledtabular}
\end{table}

In Regge analyses of the 
hadron spectrum
it is customary to consider as the $I^\eta=\frac{1}{2}^+$ 
parent trajectory the one
containing the states 
in Table~\ref{tab:regge12p}
and higher spins if available.
This trajectory contains two nearly degenerate 
Regge trajectories corresponding 
to odd and even signatures. 
The degeneracy appears when the exchange 
forces are weak and, then, 
both trajectories overlap~\cite{Collins:1977jy}.
This was the case for both $\Lambda$ and $\Sigma$ trajectories 
in~\cite{Fernandez-Ramirez:2015fbq} but it is not the
case for the $\frac{1}{2}^+$ states 
as it is apparent in Fig.~\ref{fig:poles12rega},
where the degeneracy is broken and
signature $\tau=+$
(the nucleon trajectory with 
$N(939)$, $N(1680)$, and $N(2220)$ states) 
and $\tau=-$
($N(1520)$ and $N(2190)$ states) trajectories
have different parameters.
In particular,
from Fig.~\ref{fig:poles12rega} 
it is apparent that $\alpha_0$ has to be different
for each signature.
Hence, we treat both trajectories separately.
We expect both fits 
to share approximately the same slope parameter 
$\alpha'$~\cite{Collins:1977jy}
and a different $\alpha_0$ that encodes information 
on the breaking of the degeneracy, 
{\it i.e.} on the exchange forces.

The inspection of the natural parity poles in
Figs.~\ref{fig:poles12rega} and~\ref{fig:poles12regb}
highlights the agreements and
disagreements among the pole extractions.
All the extractions reasonably agree
for $\Re[s_p]$ for all the states
poles but either disagree or have very large 
uncertainties for $N(2190)$ and $N(2220)$ widths.
We note how BnGa and SAID(SE) extractions of
$N(2190)$ separate from the expected straight line
depicted in Fig.~\ref{fig:poles12rega}. 
This is interesting because 
$I^\eta_{(\tau)}=\frac{1}{2}^+_{(+)}$
and $\frac{1}{2}^+_{(-)}$ trajectories are expected to 
have the same slope $\alpha'$~\cite{Collins:1977jy}, 
and the position
of $N(2190)$ for both extractions 
is at odds with this expectation.
Considering both 
Figs.~\ref{fig:poles12rega} and~\ref{fig:poles12regb},
only J\"uBo and CMB provide a $N(2190)$ 
extraction that conforms to the
expected position of the pole within uncertainties,
although the CMB error is very large.
For $N(2220)$ all the analyses 
coincide on $\Re\left[ s_p\right]$ but differ wildly
regarding the 
width.\footnote{We remind the reader that the deeper 
in the complex plane the pole is, the 
larger the systematic uncertainties associated 
to the models and to the analytic continuation
into the unphysical Riemann sheets.}

\begin{figure}
\rotatebox{0}{\scalebox{0.31}[0.31]{\includegraphics{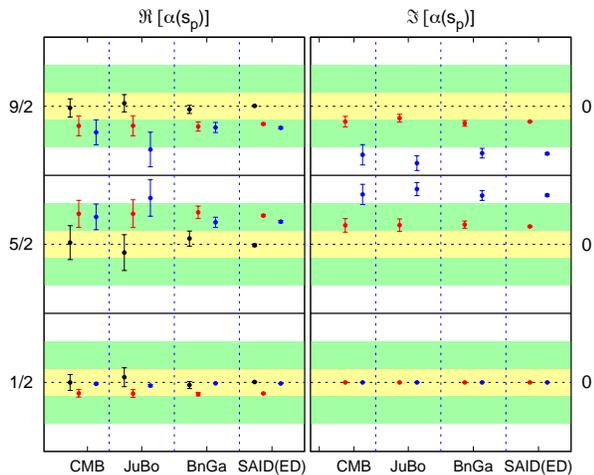}}} 
\caption{
Consistency checks 
(see Sec.~\ref{sec:fits}) 
for $I^\eta_{(\tau)}=\frac{1}{2}^+_{(+)}$
poles from CMB, J\"uBo, BnGa, and SAID(ED) extractions.
The left plot shows 
$\Re \left[ \alpha(s_p) \right]$ 
(see Table~\ref{tab:regge12p} and 
Sec.~\ref{sec:12pregge_trajectory} 
for their definition),
computed at the poles of the resonances ($s_p$) 
for models 0 (black), I (red) and II (blue). 
The result should be equal to the corresponding 
angular momentum $\Re[J]=J_p$ (vertical axis)
for a given resonance. 
The right plots depict the same calculation
for $\Im \left[ \alpha(s_p)\right]$, 
which should be equal to $\Im[J]=\Im[J_p]=0$. 
In this latter case we do not show model 0 because
$\Im \left[ \alpha(s_p)\right]=0$ by definition. 
The yellow (green) bands represent up to $0.1$ 
(from $0.1$ to $0.3$) deviation from the label 
in the vertical axis. The white band represents from $0.3$
to $0.5$ deviation.} 
\label{fig:consistency12}
\end{figure}
The comparison between 
our fitted Regge trajectories and the resonances {\it at} 
the pole positions $s_p$
($\alpha(s)$ {\it vs.} $J_p$)
are provided by the
consistency checks as described  
in~\cite{Fernandez-Ramirez:2015fbq}.
Specifically, once we have the fit parameters 
we can use them to compute the value of the Regge trajectory at the 
pole positions, hence, for a resonance 
with pole position $s_p$ 
and spin $J_p$ we should recover
$\Re\left[ \alpha(s_p)\right]=\Re [J]=J_p$ 
and $\Im \left[\alpha(s_p)\right]=\Im [J]=\Im [J_p]=0$.
This provides a direct comparison of $\alpha(s)$ (both real and imaginary parts)
to the poles, and better assesses visually the quality of the fit by
comparing the fit to $J_p$ at the poles.
The $\Im \left[\alpha(s_p)\right]=0$ condition is particularly stringent.
Moreover, the consistency check plot constitutes the appropriate
figure to compare the fit results to the fitted poles.
Consistency checks for trajectories with only two poles
do not provide any information because they are overfitted, 
(four experimental points, two masses and two widths, fitted with 
four parameters). 
Hence we only compute the consistency checks 
for trajectories with more than
two poles. The uncertainties in the poles and the parameters 
are propagated to the calculation of $\alpha(s)$.

Figure~\ref{fig:consistency12} shows the 
consistency checks for $\frac{1}{2}^+_{(+)}$ for
CMB, J\"uBo, BnGa and SAID(ED)
which provide a sharper comparison.
The consistency checks for SAID(SE), 
KH80 and KA84 are redundant and we do not show them. 
The $\frac{1}{2}^+_{(-)}$ consistency checks 
are not shown because they are overfitted and 
do not provide any information.
The $\frac{1}{2}^+_{(+)}$ does provide insight,
showing how the poles deviate from the proposed model.
If we ignore model 0, which misses the resonant physics,
the nondispersive model (I) provides, 
on average, a better consistency check than
the dispersive one (II) for all the extractions.
However, this better description of 
$N(1680)\: J_p^P=5/2^+$ and 
$N(2220)\: 9/2^+$ states is achieved by
spoiling the agreement with the nucleon $N(939)\: 1/2^+$.
These are clear indications that there is tension between
the states and our trajectory parametrization.
The $N(2220)$ has large uncertainties for all 
the extractions and its weight
on the determination of the Regge trajectory is smaller than the
nucleon and the $N(1680)$ states, which have small errors.
Besides, all the extractions agree fairly well 
regarding the pole position
of the $N(1680)$.
Hence, there is a strong indication that the approximation
of $\gamma \phi(s,s_t)=\rho(s,s_t)$ is not valid for the $N(1680)$,
signaling a sizeable contribution from physics beyond
the compact $3q$ picture.
We note that constitutent quark models 
have problems reproducing the mass of this
state and they usually overestimate
it~\cite{Capstick:1986bm,Bijker:2000gq,Loring:2001kx}.

These differences are more apparent 
if we compare the fits
to the pole sets
with the three models.
We provide the fit parameters in
Tables~\ref{tab:nat12palpha0}--\ref{tab:nat12pg}.
First, the value of $s_t$ represents
an effective threshold for the phase space
and its fitted value should be consistent with
such interpretation,
{\it i.e.} 
$s_t \sim (m_\pi+m_N)^2 \simeq 1.17\: \text{GeV}^2$.
This is used as a criterion to select the
physically meaningful minimum if several 
local minima appear in the fits, and 
to partly assess the quality of the Regge parameters.
For the $\frac{1}{2}^+_{(+)}$ trajectory, all $s_t$ 
in Table~\ref{tab:nat12pg} are
reasonable for model II
(between  $0.92$ and $1.3$ GeV$^2$) while
they are larger for model I
(between  $1.8$ and $2.65$ GeV$^2$.)
This asserts the better physical motivation of model II
compared to model I.
Therefore, we consider the parameters provided by model II
as more reliable.
For $\frac{1}{2}^+_{(-)}$ we only have two states
to estimate the trajectory parameters,
however it is enough to test, 
together with the information on $\frac{1}{2}^+_{(+)}$, 
how well the states conform to the
$\gamma \phi(s,s_t)=\rho(s,s_t)$ hypothesis.
Both models provide a large value for $s_t$ 
ranging from $2.17$ to $2.9$,
hence the slope extraction is not as reliable as for the 
$\frac{1}{2}^+_{(+)}$ trajectory.

The slope parameter $\alpha'$ links low-lying resonances and 
high-energy scattering physics, 
{\it e.g.} nucleon-antinucleon annihilation,
as it drives the Reggeon exchange amplitude under the 
single pole exchange approximation~\cite{Collins:1977jy}.
Its value is usually taken from linear 
fits to the Chew-Frautschi plot using model 0
or estimated from proton-antiproton scattering as
$\alpha' \simeq 0.98 \:
\text{GeV}^{-2}$~\cite{VandeWiele:2010kz}.
For $\frac{1}{2}^+_{(+)}$ 
we find that the $\alpha'$ extraction is
very consistent across the pole extractions.
Restricting ourselves to model II, we can estimate the slope
as 
\begin{equation}
\alpha'_{\frac{1}{2}^+_{(+)}} = 0.99\pm 0.12 \,
\text{GeV}^{-2},\nonumber
\end{equation}
where the best value and 
the uncertainty have been computed
averaging through a bootstrap
the seven $\alpha'$ in Table~\ref{tab:nat12palphaprime}.
These values are not very different 
from the ones obtained with model 0,
$\alpha'^{(0)}\simeq 1\, \text{GeV}^{-2}$,
and neglecting the widths 
does not have a large impact in $\alpha'$.
These results are also
in agreement with what is expected from
algebraic~\cite{Bijker:2000gq,Ortiz-Pacheco:2018ccl}  
($\alpha'=1.07\pm 0.02\, \text{GeV}^{-2}$)
and relativistic~\cite{Loring:2001kx} 
($\alpha'\simeq 1\,\text{GeV}^{-2}$)
quark models,
despite the fact that they 
miss dynamics~\cite{FernandezRamirez:2007ux}
that are present in the actual Regge trajectories.
The $\frac{1}{2}^\pm$ trajectories
should have the same 
slope~\cite{Collins:1977jy}, hence
once we have a robust determination from the 
$\frac{1}{2}^+_{(+)}$ we can use it to 
benchmark and assess the parameters extracted
from other trajectories.

Regarding the $\frac{1}{2}^+_{(-)}$ slope,
all pole extractions agree for model I 
and are consistent with $\frac{1}{2}^+_{(+)}$.
However, we find large differences for model II.
The only extractions that provide a consistent 
picture throughout the three models of the trajectory
are BnGa and J\"uBo,
{\it i.e.} 
$\sqrt{s_t} \simeq  1.45 - 1.55\: \text{GeV}$ 
is closer to the expected value 
of $\sqrt{s_t} \sim m_\pi +m_p \simeq 1.08\: \text{GeV}$
than the other  pole sets and $\alpha' \sim 1$ GeV$^{-2}$
close to the extracted value from 
$\frac{1}{2}^+_{(+)}$ trajectory.
Although J\"uBo has model II 
slope slightly larger than expected.
The $N(1520)$ state is very well established 
and all the pole extractions agree.
Hence, a better knowledge of this trajectory 
and an assessment on the nature of its states
based on Regge phenomenology
requires a better determination
of the $N(2190)$ state and the $N\: 11/2^-$ state.
\begin{table}
\caption{$\Delta\alpha_0 \equiv 
\alpha_0(\tau=+)-\alpha_0(\tau=-)$
for the $\frac{1}{2}^+$ trajectories and the three models.
Uncertainties obtained adding errors in quadrature.}
\label{tab:alpha0diff}
\begin{ruledtabular}
\begin{tabular}{lccc}
Pole set&Model 0&Model I &Model II\\
\hline
CMB &$\phantom{-}0.2(1)$ &$1.1(4)$& $3.8(8)$\\
J\"uBo &$\phantom{-}0.4(1)$&$1.4(1)$&$2.4(3)$\\
BnGa & $-0.02(9)$&$0.7(1)$& $1.6(5)$\\
SAID(SE) &$\phantom{-}0.11(7)$ & $1.2(1)$& $4.8(3)$\\
SAID(ED) &$\phantom{-}0.45(4)$ &$1.54(6)$& $5.84(7)$
\end{tabular}
\end{ruledtabular}
\end{table}

As expected, $\alpha_0$ is different 
for the two signatures (Table~\ref{tab:nat12palpha0}). 
Considering $\frac{1}{2}^+_{(+)}$, 
the values of $\alpha_0$ are very 
similar for models I and II across the different pole sets
and different from model 0. 
Here we appreciate the impact in the
trajectory parameter extraction 
due to the inclusion of the resonant nature of the states.
However, the values of $\alpha_0$ for $\frac{1}{2}^+_{(-)}$
change a lot from model to model 
and from pole extraction to pole extraction.
This is mostly due to the discrepancies 
among models in the extraction of
the width of $N(2190)$.
In Table~\ref{tab:alpha0diff}
we provide the difference
\makebox{$\Delta \alpha_0=\alpha_0\left( \tau=+ \right)
-\alpha_0\left( \tau=- \right)$},
for each model and pole extraction
as a way to quantify the degeneracy breaking.
The fact that each amplitude analysis 
provides a different value
for $\Delta \alpha_0$ shows that the strength 
of the exchange forces are different among them.
These forces are related to the left-hand cut of the amplitudes
and are not well known.
Hence, the range of values for $\Delta \alpha_0$ quantifies 
the magnitude of the uncertainties associated 
to this particular model dependency.
Inspecting Table~\ref{tab:alpha0diff} it is noticeable that
$\Delta \alpha_0$ for BnGa and model 0 is negative. 
This is related to the difference 
in the extraction of the slope parameter
$\alpha'$ ($1.07(3)$ and $0.85(3)$ in
Table~\ref{tab:nat12palphaprime}).
However, if we introduce the widths in the analysis, 
$\Delta \alpha_0$
becomes positive (as expected 
from Fig.~\ref{fig:poles12rega})
and 
the slopes become compatible within errors
($0.87(3)$ and $0.86(3)$ for model I 
and $1.04(6)$ and $1.15(6)$ for model II).
This again shows the importance of including 
the width in the analysis, and, moreover,
how its inclusion leads to a better 
and more consistent estimation of both $\alpha_0$
and the slope parameter $\alpha'$. 
Our best estimation of $\alpha_0$, 
using the same technique as for $\alpha'$
and model II, is
\begin{equation}
\alpha_{0,\frac{1}{2}^+_{(+)}} = 0.21\pm 0.38\, .\nonumber
\end{equation}
The two remaining parameters are
\begin{equation}
\gamma_{\frac{1}{2}^+_{(+)}} = 0.651 \pm 0.040; \quad
s_{t,\frac{1}{2}^+_{(+)}}=1.02 \pm 0.13\, \text{GeV}^2, 
\nonumber
\end{equation}
with the effective threshold close to
the expected value of
$(m_\pi + m_p)^2\simeq 1.17$~GeV$^2$.

\subsubsection{$\frac{1}{2}^-$ Regge trajectory}
\label{sec:12mregge_trajectory}
\begin{table}
\caption{Parameter $\alpha_0$ 
obtained for $\frac{1}{2}^-$ trajectories.} 
\label{tab:nat12malpha0}
\begin{ruledtabular}
\begin{tabular}{clccc}
$I^\eta_{(\tau)}$&Pole set&
$\alpha_0^{(0)}$ &$\alpha_0^{(\text{I})}$ & $\alpha_0^{(\text{II})}$\\
\hline
$\frac{1}{2}^-_{(+)}$ & CMB &$-0.4(3)\phantom{0}$&$-0.7(3)\phantom{0}$&$-3(2)\phantom{.00}$\\
&J\"uBo&$-4(1)\phantom{.00}$&$-4(1)\phantom{.00}$&$-7(3)\phantom{.00}$\\
 & BnGa&$-0.1(2)\phantom{0}$&$-0.5(2)\phantom{0}$&$-6(1)\phantom{.00}$\\
&SAID(SE)&$\phantom{-}0.25(3)$&$\phantom{-}0.16(4)$&$-0.5(2)\phantom{0}$\\
& SAID(ED)&$\phantom{-}0.01(3)$&$-0.21(3)$&$-2.3(1)\phantom{0}$\\
& KH80 &$-0.4(1)\phantom{0}$&$-0.6(1)\phantom{0}$&$-4(1)\phantom{.00}$\\
&KA84 &$-0.19(3)$&$-0.41(5)$&$-3.0(5)\phantom{0}$\\
\hline
$\frac{1}{2}^{-}_{(-)}$ & CMB &$-6(1)\phantom{.00}$&$-6(2)\phantom{0.0}$&$-9(2)\phantom{.00}$\\
&J\"uBo &$-1.7(1)\phantom{0}$&$-1.8(1)\phantom{0}$&$-2.1(1)\phantom{0}$\\
 & BnGa &$-3.6(5)\phantom{0}$&$-3.0(6)\phantom{0}$&$-3.0(6)\phantom{0}$\\
&SAID(SE) &$-1.5(4)\phantom{0}$&$-1.5(4)\phantom{0}$&$-0.38(3)\phantom{0}$\\
& KH80 &$-2.2(1)\phantom{0}$&$-2.9(2)\phantom{0}$&$-10.2(4)$\\
& KA84 &$-2.5(1)\phantom{0}$&$-3.2(2)\phantom{0}$&$-11.2(4)$\\
\end{tabular}
\end{ruledtabular}
\end{table}
\begin{table}
\caption{Parameter $\alpha'$ obtained for 
$\frac{1}{2}^-$ trajectories.} 
\label{tab:nat12malphaprime}
\begin{ruledtabular}
\begin{tabular}{clccc}
$I^\eta_{(\tau)}$ &Pole set&$\alpha'^{(0)}$ &$\alpha'^{(\text{I})}$ & $\alpha'^{(\text{II})}$\\
\hline
$\frac{1}{2}^-_{(+)}$ & CMB &$1.1(1)\phantom{0}$&$1.1(1)$&$1.8(5)\phantom{0}$\\
& J\"uBo &$2.3(5)\phantom{0}$&$2.3(5)\phantom{0}$&$3(1)\phantom{.00}$\\
& BnGa &$0.97(7)$&$0.99(7)$&$2.1(2)\phantom{0}$\\
& SAID(SE) &$0.81(1)$&$0.82(1)$&$1.03(4)$\\
& SAID(ED) &$0.91(1)$&$0.93(1)$&$1.46(2)$\\
& KH80 &$1.04(3)$&$1.07(4)$&$1.8(2)\phantom{0}$\\
& KA84 &$0.98(1)$&$0.99(1)$&$1.6(1)\phantom{0}$\\
\hline
$\frac{1}{2}^-_{(-)}$ & CMB &$2.6(4)\phantom{0}$&$2.6(4)\phantom{0}$&$3.4(5)\phantom{0}$\\
&J\"uBo &$1.13(3)$&$1.13(3)$&$1.28(4)$\\
& BnGa &$1.8(2)\phantom{0}$&$1.6(2)\phantom{0}$&$1.9(2)\phantom{0}$\\
&SAID(SE) &$1.1(1)\phantom{0}$&$1.1(1)\phantom{0}$&$1.18(1)$\\
& KH80 &$1.32(4)$&$1.37(4)$&$3.2(1)\phantom{0}$\\
& KA84 &$1.40(5)$&$1.50(5)$&$3.5(1)\phantom{0}$\\
\end{tabular}
\end{ruledtabular}
\end{table}
\begin{table}
\caption{Parameters $\gamma$ and $s_t$
obtained for $\frac{1}{2}^-$ trajectories.}
\label{tab:nat12mg}
\begin{ruledtabular}
\begin{tabular}{clccccc}
$I^\eta_{(\tau)}$&Pole set&$\gamma^{(\text{I})}$&  $\gamma^{(\text{II})}$&$s_t^{(\text{I})}$&$s_t^{(\text{II})}$\\
\hline
$\frac{1}{2}^-_{(+)}$&CMB&$0.6(2)\phantom{0}$&$3(1)\phantom{.00}$&$2.6(2)\phantom{0}$&$3.0(3)\phantom{0}$\\
&J\"uBo&$1.0(4)\phantom{0}$&$2(1)\phantom{.00}$&$2.2(5)\phantom{0}$&$2.5(4)\phantom{0}$\\
& BnGa &$0.70(9)$&$3.2(4)\phantom{0}$&$2.73(4)$&$3.4(1)\phantom{0}$\\
&SAID(SE) &$0.34(3)$&$0.9(1)\phantom{0}$&$2.44(7)$&$2.7(1)\phantom{0}$\\
& SAID(ED) &$0.56(1)$&$1.84(5)$&$2.69(2)$&$3.07(2)$\\
& KH80 &$0.67(8)$&$1.14(5)$&$2.72(4)$&$3.1(1)\phantom{0}$\\
&KA84 &$0.59(3)$&$1.8(2)\phantom{0}$&$2.71(2)$&$3.0(1)\phantom{0}$\\
\hline
$\frac{1}{2}^-_{(-)}$ &CMB &$1.4(5)\phantom{0}$&$3(1)\phantom{0.0}$&$2.6(4)\phantom{0}$&$2.7(3)\phantom{0}$\\
&J\"uBo &$0.31(4)$&$0.8(1)\phantom{0}$&$1.3(4)\phantom{0}$&$2.3(1)\phantom{0}$\\
& BnGa &$0.6(1)\phantom{0}$&$1.3(1)\phantom{0}$&$1.02(4)$&$1.1(1)\phantom{0}$\\
&SAID(SE) &$0.3(1)\phantom{0}$&$0.63(2)\phantom{0}$&$0.8(1)\phantom{0}$&$1.52(1)$\\
& KH80 &$1.2(1)\phantom{0}$&$5.0(2)\phantom{0}$&$2.84(3)$&$3.31(3)$\\
& KA84 &$1.3$(1)\phantom{0}&$5.5(2)\phantom{0}$&$2.92(2)$&$3.31(2)$\\
\end{tabular}
\end{ruledtabular}
\end{table}

In Table~\ref{tab:regge12m} 
we provide the lowest-lying states for each spin $J_p$
compatible with the $\frac{1}{2}^-$ Regge trajectory
except for the $N(1535)$ ($J^P_p=1/2^-$) 
which belongs to a daughter trajectory~\cite{Collins:1977jy}.
As for $\frac{1}{2}^+$ trajectory, 
we have two nearly degenerate trajectories
with opposite signatures.
However, the $(\Im \left[ s_p \right],J_p)$ 
plot in Fig.~\ref{fig:poles12regb}
provides conflicting information 
about the $N(1720)\, 3/2^+$ state.
The large widths obtained by
BnGa, SAID(SE) and SAID(ED),
$\Gamma_p \sim 300 - 430$~MeV,
would place this state in the daughter trajectory.
However, CMB is compatible with $N(1720)$ 
($\Gamma_p=120$ MeV)
belonging to the parent trajectory,
and J\"uBo, KH80, and KA84 
($\Gamma_p \sim 185$~MeV)
are in between both possibilities.
If we look into the other pole extractions that 
we do not consider in our analysis, we
see that
SAID obtains $334$~MeV~\cite{Svarc:2014aga},
similar to BnGa, SAID(SE) and SAID(ED).
Other pole sets are closer to the 
J\"uBo, KH80 and KA84 extractions,
{\it e.g.} 
H\"ohler $187$~MeV~\cite{Hoehler1993},
KSU $175$~MeV~\cite{Shrestha:2012ep},
and Zagreb $233$~MeV~\cite{Batinic:2010zz};
while others obtain smaller 
widths compatible with the CMB result
{\it e.g.}
P-ANL $94$ MeV~\cite{Vrana:1999nt},
Giessen $118$~MeV~\cite{Shklyar:2012js},
and ANL-O $70$ MeV~\cite{Kamano:2013iva}.
We note that the discrepancies among pole extractions,
together with constituent quark models 
predicting several $3/2^+$ states in the $N(1720)$ energy
range~\cite{Capstick:1986bm,Bijker:1994yr,Bijker:2000gq,Loring:2001kx}
make possible that what the different amplitude
analysis are reporting is not just one resonant state
but an effective pole that accounts 
for a more complicated picture.
Moreover, the recent
ANL-O pole extraction finds two states
with masses $1703$ and $1763$~MeV
and widths $70$ and $159$~MeV
respectively~\cite{Kamano:2013iva}.
Further research on this energy range is necessary
to establish mass and width of the state(s) with precision
before discussing its (their) nature.
In what follows, we include $N(1720)$ in our calculations as a 
member of the parent $\frac{1}{2}^-_{(-)}$ trajectory.

Contrary to $\frac{1}{2}^+$ resonances, 
$\frac{1}{2}^-$ states that belong to the
leading Regge trajectory are not that well known, 
what predates any conclusion
on the internal structure of the states
that we can derive from fits.
At this stage, Regge phenomenology can be used more effectively
as a guide to improve
amplitude analyses and pole extraction
than to elucidate the nature of the
resonances.

Figures~\ref{fig:polesrega} and~\ref{fig:polesregb}
make apparent how different
are the poles from one extraction to another.
There is consensus only on the $N(1675)\, 5/2^-$ state.
This is a direct challenge 
to the four-star status of $N(1720)$ 
and $N(2250)$ resonances in the PDG~\cite{Tanabashi:2018oca}.
We fit two trajectories 
$\frac{1}{2}^-_{(+)}$ ($N(1675)$ and $N(2250)$ states)
and $\frac{1}{2}^-_{(-)}$ ($N(1720)$
and $N(1990)$ states).
The obtained fit parameters are provided in 
Tables~\ref{tab:nat12malpha0}--\ref{tab:nat12mg}.
For the $\frac{1}{2}^-_{(+)}$, none of the 
pole extractions provides a good result for $s_t$.
Besides, MacDowell 
symmetry~\cite{Collins:1977jy,MacDowell:1959zza}
imposes that the slopes for
$\frac{1}{2}^+_{(+)}$ and $\frac{1}{2}^-_{(-)}$ 
($\frac{1}{2}^+_{(-)}$ and $\frac{1}{2}^-_{(+)}$)
should be equal. Hence, 
we should obtain $\alpha' \sim 1$~GeV$^{-2}$ to agree with
the results in Sec.~\ref{sec:12pregge_trajectory},
a condition only SAID(SE) fulfills for the three models,
despite the fact that its $s_t=2.7$~GeV$^2$ 
is larger than expected.
Regarding negative signature, 
only BnGa and SAID(ED) are close to $s_t\sim 1.2$~GeV$^2$.
If we also consider the expected slope, 
the only pole extraction
that provides reasonable parameters is SAID(ED).
Finally, J\"uBo provides a higher $s_t =2.3$ 
and a slightly large
but reasonable slope.
We do not provide plots with the consistency check 
as both trajectories are overfitted.

In summary, none of the pole sets provides 
a convincing picture of the $\frac{1}{2}^-$
trajectory and there is a reasonable possibility 
that $N(1720)$
actually belongs to the parent trajectory.
This state is a doublet partner of the $N(1680)$, which 
we identified in Sec.~\ref{sec:12pregge_trajectory}
as a state with physics beyond the compact $3q$ picture.
This makes $N(1720)$ a prime candidate 
to look for additional dynamics, and
explains why it might be displaced 
from the expected pattern and can be
missidentified as a member of a daughter trajectory.
This state also shows how the inclusion 
of the width and the patterns in the 
$(\Im[s_p],J_p)$ allows to better identify 
if a state is in the leading
trajectory or in a subleading one.
Again, a better determination of this state 
would allow further investigation
on its nature.

\subsubsection{$\frac{3}{2}^+$ Regge trajectory}
\label{sec:32pregge_trajectory}
\begin{table}
\caption{Parameter $\alpha_0$ obtained for $\frac{3}{2}^+$ trajectory.} 
\label{tab:nat32palpha0}
\begin{ruledtabular}
\begin{tabular}{clcccc}
$I^\eta_{(\tau)}$&Pole set&$\alpha_0^{(0)}$ &$\alpha_0^{(\text{I})}$ & $\alpha_0^{(\text{II})}$\\
\hline
$\frac{3}{2}^+_{\phantom{(+)}}$
& CMB &$-1.2(4)$&$-1.3(4)$&$-1.6(6)\phantom{0}$\\
& J\"uBo &$-1.3(2)$&$-1.0(2)$&$-1.0(3)\phantom{0}$\\
& BnGa &$-5.7(6)$&$-5.7(6)$&$-6.0(8)\phantom{0}$\\
& SAID(SE) &$-2.7(3)$&$-3.2(3)$&$-7(1)\phantom{.00}$\\
& SAID(ED) &$-3.4(3)$&$-3.5(3)$&$-4.5(6)\phantom{0}$\\
& KH80 &$-5.9(6)$&$-7.2(8)$&$-22.7(2)$\\
& KA84 &$-2.9(2)$&$-3.0(2)$&$-3.5(1)\phantom{0}$\\
\hline
$\frac{3}{2}^+_{(+)}$
&CMB &$-0.5(5)$&$-0.5(4)$&$-1.2(6)\phantom{0}$\\
\hline
$\frac{3}{2}^+_{(-)}$
&CMB &$-2.1(4)$&$-2.2(5)$&$-4(1)\phantom{.00}$\\
&J\"ubo &$\phantom{-}1.0(7)$&$-1.2(6)$&$-1.8(9)\phantom{0}$\\
\end{tabular}
\end{ruledtabular}
\end{table}
\begin{table}
\caption{Parameter $\alpha'$ obtained for $\frac{3}{2}^+$ trajectory.} 
\label{tab:nat32palphaprime}
\begin{ruledtabular}
\begin{tabular}{clcccc}
$I^\eta_{(\tau)}$&Pole set&$\alpha'^{(0)}$ &$\alpha'^{(\text{I})}$ & $\alpha'^{(\text{II})}$\\
\hline
$\frac{3}{2}^+_{\phantom{(+)}}$
& CMB &$1.0(1)$&$1.0(1)\phantom{0}$&$1.2(2)\phantom{0}$\\
& J\"uBo &$1.0(1)$&$1.01(4)$&$1.0(1)\phantom{0}$\\
& BnGa &$2.5(2)$&$2.5(2)\phantom{0}$&$2.7(3)\phantom{0}$\\
& SAID(SE) &$1.6(1)$&$1.6(1)\phantom{0}$&$1.38(8)$\\
& SAID(ED) &$1.8(1)$&$1.8(1)\phantom{0}$&$2.2(2)\phantom{0}$\\
& KH80 &$2.7(2)$&$2.9(2)\phantom{0}$&$7.6(1)\phantom{0}$\\
& KA84 &$1.7(1)$&$1.7(1)\phantom{0}$&$2.00(2)$\\
\hline
$\frac{3}{2}^+_{(+)}$
&CMB &$0.9(1)$&$0.9(1)\phantom{0}$&$1.1(2)\phantom{0}$\\
\hline
$\frac{3}{2}^+_{(-)}$
&CMB &$1.3(1)$&$1.3(1)\phantom{0}$&$1.9(4)\phantom{0}$\\
&J\"ubo &$0.9(2)$&$0.9(2)\phantom{0}$&$1.1(3)\phantom{0}$\\
\end{tabular}
\end{ruledtabular}
\end{table}
\begin{table}
\caption{Parameters $\gamma$ and $s_t$ obtained for $\frac{3}{2}^+$ trajectory.}
\label{tab:nat32pg}
\begin{ruledtabular}
\begin{tabular}{clcccc}
$I^\eta_{(\tau)}$&Pole set&$\gamma^{(\text{I})}$&  $\gamma^{(\text{II})}$&$s_t^{(\text{I})}$&  $s_t^{(\text{II})}$\\
\hline
$\frac{3}{2}^+_{\phantom{(+)}}$
& CMB &$0.5(1)$&$1.2(3)$&$2.0(6)$&$2.3(4)\phantom{0}$\\
& J\"uBo&$0.5(2)$&$1.1(3)$&$2.0(2)$&$2.5(4)\phantom{0}$\\
& BnGa&$0.9(1)$&$1.8(3)$&$0.9(2)$&$1.4(5)\phantom{0}$\\
& SAID(SE) &$1.0(1)$&$3.0(5)$&$2.5(1)$&$2.8(1)\phantom{0}$\\
& SAID(ED) &$0.7(1)$&$1.6(3)$&$1.3(7)$&$2.1(4)\phantom{0}$\\
& KH80 &$2.4(3)$&$9.4(1)$&$2.7(1)$&$2.94(2)$\\
& KA84 &$0.6(1)$&$1.4(1)$&$0.8(5)$&$1.8(2)\phantom{0}$\\
\hline
$\frac{3}{2}^+_{(+)}$
&CMB &$0.4(1)$&$1.3(4)$&$1.7(5)$&$2.7(4)\phantom{0}$\\
\hline
$\frac{3}{2}^+_{(-)}$
&CMB &$0.6(2)$&$2.0(1)$&$1.9(6)$&$2.7(3)\phantom{0}$\\
&J\"ubo &$0.6(3)$&$1.3(6)$&$2.5(1)$&$2.5(3)\phantom{0}$\\
\end{tabular}
\end{ruledtabular}
\end{table}

This is the least known parent trajectory, with
two well established states
--$\Delta(1700)$ and $\Delta(1905)$--
and  only CMB and J\"uBo reporting
additional resonances. 
Hence, not much information can be obtained from this
trajectory.
Comparing all the extractions for 
$\Delta(1700)$ and $\Delta(1905)$ we see
in Figs.~\ref{fig:poles32rega} and \ref{fig:poles32regb} that
$\Re [s_p]$ is reasonably established 
for both but the width presents large uncertainties.
If we consider the CMB and J\"uBo $7/2^-$ state
and CMB $9/2^+$
in Fig.~\ref{fig:poles32rega} 
a degeneracy breaking is hinted.
Hence, we first 
fit the $\frac{3}{2}^+$ trajectory without
considering the degeneracy breaking
for all the pole extractions and we also fit
$\frac{3}{2}^+_{(+)}$ for J\"uBo and 
$\frac{3}{2}^+_{(\pm)}$ for CMB.
We provide the parameters in 
Tables~\ref{tab:nat32palpha0}--\ref{tab:nat32pg}.
Because we assume degeneracy in $\frac{3}{2}^+$ fits,
the $\alpha_0$ parameter provides no information.
Also, the value of $s_t$ is highly correlated with
$\alpha_0$, so it is not possible to use its value 
as a way to assess the quality of the extracted 
parameters.
It is clear that degeneracy is a bad approximation to
obtain the Regge parameters.
Hence, we do not provide consistency checks 
for this trajectory as they do not provide
insight.
We note that CMB and J\"uBo provide a reasonable
slope $\alpha'\simeq 1$~GeV$^{-2}$.
J\"uBo (CMB) provides a consistent slope parameter
for $\frac{3}{2}^-_{(-)}$ ($\frac{3}{2}^-_{(+)}$)
once degeneracy breaking is considered with 
$\alpha'\simeq 1$~GeV$^{-2}$.
However, CMB provides a very large 
slope for $\frac{3}{2}^-_{(-)}$.
The overall picture, makes the 
J\"uBo extraction of $\frac{3}{2}^+$ the most consistent
one, although with very large error bars.

\subsubsection{$\frac{3}{2}^-$ Regge trajectory}
\label{sec:32mregge_trajectory}
\begin{table}
\caption{Parameter $\alpha_0$ obtained for 
$\frac{3}{2}^-$ trajectories.} 
\label{tab:nat32malpha0}
\begin{ruledtabular}
\begin{tabular}{clcccc}
$I^\eta_{(\tau)}$&Pole set&$\alpha_0^{(0)}$ &$\alpha_0^{(\text{I})}$ & $\alpha_0^{(\text{II})}$\\
\hline
$\frac{3}{2}^-_{(+)}$
& J\"uBo  &$-8(8)$&$-11(10)$&$-9(12)$\\
& SAID(ED) &$13(9)$&$-75(1)\phantom{0}$&$34.3(8)$\\
\hline
$\frac{3}{2}^-_{(-)}$
& CMB &$\phantom{-}0.1(2)\phantom{0}$&$-0.1(4)\phantom{0}$&$-0.4(5)\phantom{0}$\\
& J\"uBo &$-0.02(8)$&$-0.1(1)\phantom{0}$&$-1.1(6)\phantom{0}$\\
& BnGa &$\phantom{-}0.10(1)$&$\phantom{-}0.05(1)$&$-0.45(4)$\\
& SAID(SE) &$\phantom{-}0.10(1)$&$\phantom{-}0.06(1)$&$-0.39(3)$\\
& SAID(ED) &$-0.03(3)$&$-0.9(3)\phantom{0}$&$-0.43(5)$\\
& KH80 &$\phantom{-}0.28(3)$&$\phantom{-}0.25(3)$&$\phantom{-}0.13(4)$\\
& KA84 &$-0.07(1)$&$-2.1(3)\phantom{0}$&$-0.51(3)$\\
\end{tabular}
\end{ruledtabular}
\end{table}
\begin{table}
\caption{Parameter $\alpha'$ obtained 
for $\frac{3}{2}^-$ trajectories.} 
\label{tab:nat32malphaprime}
\begin{ruledtabular}
\begin{tabular}{clcccc}
$I^\eta_{(\tau)}$&Pole set&$\alpha'^{(0)}$ &$\alpha'^{(\text{I})}$ & $\alpha'^{(\text{II})}$\\
\hline
$\frac{3}{2}^-_{(+)}$
& J\"uBo  &$\phantom{-}4(3)$&$4(3)\phantom{0}$&$\phantom{-}5(4)$\\
& SAID(ED) &$-3(2)$&$8.0(2)$&$-4.1(3)$\\
\hline
$\frac{3}{2}^-_{(-)}$
& CMB &$0.97(8)$&$1.0(1)\phantom{0}$&$1.2(2)$\\
& J\"uBo &$1.03(5)$&$1.04(5)$&$1.4(2)$\\
& BnGa &$0.95(1)$&$0.95(1)$&$1.19(1)$\\
& SAID(SE) &$0.953(4)$&$0.958(4)$&$1.17(1)$\\
& SAID(ED) &$1.02(1)$&$1.18(5)$&$1.23(2)$\\
& KH80 &$0.87(1)$&$0.87(1)$&$1.00(2)$\\
& KA84 &$1.04(1)$&$1.36(5)$&$1.28(1)$\\
\end{tabular}
\end{ruledtabular}
\end{table}
\begin{table}
\caption{Parameters $\gamma$ and $s_t$ obtained for $\frac{3}{2}^-$ trajectory.}
\label{tab:nat32mg}
\begin{ruledtabular}
\begin{tabular}{clccccc}
$I^\eta_{(\tau)}$&Pole set&$\gamma^{(\text{I})}$&  $\gamma^{(\text{II})}$& $s_t^{(\text{I})}$& $s_t^{(\text{II})}$\\
\hline
$\frac{3}{2}^-_{(+)}$
& J\"uBo  &$4(4)$&$4(5)$&$3(3)$&$4(5)$\\
& SAID(ED) &$28.(2)$&$-8(1)$&$6.6(1)$&$10(2)$\\
\hline
$\frac{3}{2}^-_{(-)}$
& CMB &$0.5(1)$&$0.9(2)$&$1.6(3)\phantom{0}$&$1.5(1)\phantom{0}$\\
& J\"uBo &$0.35(7)$&$0.9(3)$&$1.34(9)$&$1.7(2)\phantom{0}$\\
& BnGa &$0.29(1)$&$0.67(2)$&$1.34(1)$&$1.54(1)$\\
& SAID(SE) &$0.28(1)$&$0.63(2)$&$1.32(1)$&$1.52(1)$\\
& SAID(ED) &$0.70(9)$&$0.98(3)$&$2.8(3)$&$1.49(1)$\\
& KH80 &$0.39(3)$&$0.80(6)$&$1.39(2)$&$1.40(1)$\\
& KA84 &$1.2(1)$&$1.16(2)$&$3.5(2)$&$1.48(1)$\\
\end{tabular}
\end{ruledtabular}
\end{table}

In this trajectory there are three four-star
resonances, 
namely $\Delta(1232)$, $\Delta(1950)$ and 
$\Delta(2420)$, all of them 
with even signature.
The first two are obtained by all the pole extractions
and agree on both mass and width.
The higher mass state is found
by CMB, SAID(ED), KH80, and KA84 analyses.
SAID(ED) and KA84 agree on $\Re[s_p]$,
see Fig.~\ref{fig:poles32rega}, while
KH80 is at odds with their result.
If we look into $\Im[s_p]$, Fig.~\ref{fig:poles32regb},
SAID(ED) and KH80 disagree, 
while KA84 extraction overlaps both of them due to its
large uncertainty.
The CMB extraction of this pole
has large uncertainties too and agrees with the other
three pole sets within errors. 

We perform fits to the odd and even signatures.
The fit parameters are reported in 
Tables~\ref{tab:nat32malpha0}--\ref{tab:nat32mg}.
The parameters for $\frac{3}{2}^-_{(+)}$
are completely at odds with the Regge expectation
and the obtained $s_t$ are not
physically sensible,
\textit{i.e.} $s_t \gg (m_p+m_\pi)^2$.
The reasons are obvious if we inspect
Fig.~\ref{fig:poles32rega}, the position
of the $9/2^-$ pole obtained by J\"uBo and SAID(ED)
has a very low $\Re[s_p]$ value
given the position of $\Delta(1930)$.
Also, in the case
of SAID(ED), $\Im[s_p]$ is too large.
Hence, the position of this pole is completely unreliable,
both in mass and width, as the large uncertainties
in the J\"uBo width hint
and no further conclusions can be derived.

Regarding the $\frac{3}{2}^-_{(-)}$
(the $\Delta$ trajectory), 
the effective threshold is at odds with the expected value 
only for model I in SAID(ED) and KA84 poles. 
For the rest of 
pole sets and for model II we obtain reasonable values.
The slopes are close to unity as expected 
and only the $\alpha_0$
value shows a large variation among models and pole sets.
We can compare our Regge parameters to those used in
fits to high energy proton-antiproton annihilation,
where $\Delta$ Regge trajectory 
$\alpha_\Delta(s)=-0.37+0.98\, s$ ($s$ in GeV$^2$)
is one of the main contributions~\cite{VandeWiele:2010kz}.
We note that the slope is 
close to unity and that the $\alpha_0$
parameter agrees with the one we obtain 
for $\frac{3}{2}^-_{(-)}$ 
using model II. 
Hence, model II provides the result compatible
with the high energy information
and our most reliable determination of the parameters.
Consequently,
as we did in Sec.~\ref{sec:12pregge_trajectory}, 
we can estimate $\alpha'$ from model II
values in Table~\ref{tab:nat32malphaprime} as
\begin{equation}
\alpha'_{\frac{3}{2}^-_{(-)}} =  
1.21\pm 0.15\, \text{GeV}^2.
\nonumber
\end{equation}
We note that this slope is compatible within errors
with the one obtained from the 
$\frac{1}{2}^+_{(+)}$ trajectory
in Sec.~\ref{sec:12pregge_trajectory}.
The remaining parameters are 
\begin{align}
\alpha_{0,\frac{3}{2}^-_{(-)} } =&
\,-0.45\pm 0.44; \nonumber \\
\gamma_{\frac{3}{2}^-_{(-)}} =&
\, 0.86 \pm 0.22; \nonumber \\
s_{t,\frac{3}{2}^+_{(-)}}=&
\,  1.52 \pm 0.12\, \text{GeV}^2, 
\nonumber
\end{align}
with the effective threshold slightly above
the expected value of
$(m_\pi + m_p)^2\simeq 1.17$~GeV$^2$.

\begin{figure}
\rotatebox{0}{\scalebox{0.31}[0.31]{\includegraphics{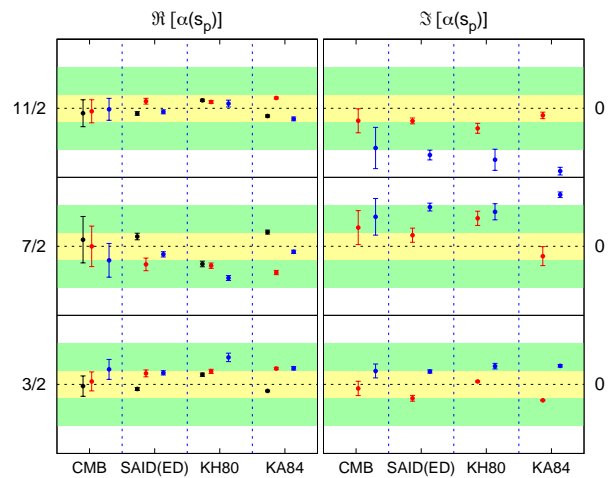}}}
\caption{Consistency checks 
for $\frac{3}{2}^-_{(-)}$
poles from 
CMB, SAID(ED), KH80 and KA84 extractions.
Notation as in Fig.~\ref{fig:consistency12}.
See Sec.~\ref{sec:32mregge_trajectory} 
for trajectory definition.} 
\label{fig:consistency32}
\end{figure} 
Figures~\ref{fig:poles32rega} 
and~\ref{fig:poles32regb} show a clear
linear and square-root-like pattern 
for the $\frac{3}{2}^-_{(-)}$
trajectory hinting that these states 
are compact $3q$ structures.
The consistency check in
Fig.~\ref{fig:consistency32} provides a sharper image.
The deviations are clear and only CMB
provides an approximate agreement between theory and data, 
mostly due to the 
large uncertainties.
Considering that CMB overlaps with the pole extractions by
other analyses, its deviation from the
trajectory models in Eq.~\eqref{eq:models}
signals the effects of beyond compact $3q$ physics,
even for the well-studied $\Delta(1232)$ state.
The $\frac{3}{2}^-_{(-)}$ 
poles are known well enough to be sensitive to these beyond compact $3q$
effects.

\section{Summary and conclusions}
\label{sec:conclusions}
We have studied the structure of the 
$N^*$ and $\Delta^*$ spectra from the perspective of
Regge and complex angular momentum theory 
following the work done for the 
strange baryon sector
in \cite{Fernandez-Ramirez:2015fbq}.
We have considered seven pole extractions
~\cite{Cutkosky:1979fy,Cutkosky:1980rh,Ronchen:2018ury,Anisovich:2011fc,Sokhoyan:2015fra,Svarc:2014aga,Svarc:2014zja}.
In our analysis we have taken
into account the fact that poles are 
complex quantities, and we go beyond the standard 
studies that focus only in the 
Chew-Frautschi plot ($\Re [s_p],J_p$) 
and linear trajectory fits to said plot.
In doing so, we also study the ($\Im [s_p],J_p$)  
plots introduced in~\cite{Fernandez-Ramirez:2015fbq}.
We find many discrepancies among the pole extractions, 
in particular for the widths,
but a clear pattern, similar to the one in the 
strange sector, appears
where the Chew-Frautschi plots follow 
the well-known approximate linear behavior,
while the ($\Im [s_p],J_p$)  plots 
show a square-root-like behavior.

Our working hypothesis has been that
the square-root-like behavior appreciated in 
Fig.~\ref{fig:polesregb}
is due to the contribution of the phase space
to the scattering amplitude~\cite{Chew:1960iv},
which is proportional to the momentum $q\sim\sqrt{s-s_t}$.
The phase space is the main contribution to
how deep in the complex plane the poles are.
Major deviations from that pattern would signal 
an important component 
of beyond compact $3q$ physics, {\it i.e.} 
additional QCD dynamics.
Under this hypothesis, a state that presents 
a linear trajectory in the 
Chew-Frautschi plot {\it and} a square-root-like behavior
would be mostly a compact $3q$ state.
Besides the qualitative analysis of the plots,
we performed a quantitative one, modeling
the Regge trajectories, fitting the poles and
cross checking the consistency of the results.
The results support the qualitative conclusions
but also signal sizable physics 
beyond the compact $3q$ picture 
for the $N(1680)$, the $N(1720)$
and some of the members of the 
$\frac{3}{2}^-_{(-)}$ trajectory. 
The last poles are known well enough
that our analysis is sensitive to beyond compact $3q$ effects.

We find that exchange degeneracy is very clearly broken in the
nonstrange sector, contrary to the strange sector.
This degeneracy breaking shows the importance of exchange forces 
in the determination 
of the low-lying nonstrange baryon spectrum.
We also find that the $\frac{1}{2}^-$ and 
$\frac{3}{2}^+$ trajectories 
are poorly known and Regge phenomenology
cannot provide insight on the internal 
structure of the baryons.
However, Regge phenomenology
serves as a guide for resonance searches. 
Particularly, 
as a way to explore if the fits to the experimental data
are improved by including resonances 
close to the expected positions
in both Chew-Frautschi
and ($\Im [s_p],J_p$)  plots.

The parameters of the $\frac{1}{2}^+_{(+)}$ (nucleon) 
and $\frac{3}{2}^-_{(-)}$ ($\Delta$)
Regge trajectories can be well established from the poles.
We estimate $\alpha'=0.99 \pm 0.12$~GeV$^{-2}$
for the nucleon trajectory
and $\alpha' =1.21 \pm 0.15$~GeV$^{-2}$
for the $\Delta$.
We note that both slopes are compatible within errors.
This range is consistent 
with $\alpha'$ obtained from fits to the
Chew-Frautschi plots, with what is predicted by
constituent quark models and with fits to 
high energy proton-antiproton annihilation.  

\begin{acknowledgments}
JASC and CFR thank Roelof Bijker for useful comments.
This work was supported by 
PAPIIT-DGAPA (UNAM, Mexico) under grants No.~IA101717
and~IA101819,
CONACYT (Mexico) under grants No.~251817 and No.~619970,
the U.S.~Department of Energy under grants 
No.~DE-AC05-06OR23177
and No.~DE-FG02-87ER40365,
Research Foundation -- Flanders (FWO),
U.S.~National Science Foundation under award numbers 
PHY-1415459 and PHY-1513524,
Ministerio de Ciencia, Innovaci\'on y Universidades (Spain) 
grant No.~FPA2016-77313-P,
and Deutsche Forschungsgemeinschaft (DFG)
through the Collaborative Research Center 
[The Low-Energy Frontier of the Standard Model (SFB 1044)] 
and the Cluster of Excellence 
[Precision Physics, Fundamental Interactions and Structure of Matter (PRISMA)].
\end{acknowledgments}
\bibliographystyle{apsrev4-1}
\bibliography{bibliography.bib}
\end{document}